\documentclass[%
 aip,
 amsmath,amssymb,
 reprint,%
]{revtex4-1}

\usepackage{graphicx}
\usepackage{dcolumn}
\usepackage{bm}

\usepackage[utf8]{inputenc}
\usepackage[T1]{fontenc}
\usepackage{mathptmx}
\usepackage{etoolbox}
\usepackage{subfigure} 
\usepackage{algorithm}
\usepackage{xcolor}
\usepackage{soul} 
\sethlcolor{yellow}
\usepackage[english]{babel}
\makeatletter
\def\@email#1#2{%
 \endgroup
 \patchcmd{\titleblock@produce}
  {\frontmatter@RRAPformat}
  {\frontmatter@RRAPformat{\produce@RRAP{*#1\href{mailto:#2}{#2}}}\frontmatter@RRAPformat}
  {}{}
}%
\makeatother

\begin{document}

\preprint{AIP/123-QED}

\title[]{Machine Learning-integrated Multiscale Simulation Framework: Bridging Scales in Associative Polymer-Colloid Suspensions}
\author{Jalal Abdolahi}

\author{Dominic Robe}%
\affiliation{ 
Soft Matter Informatics Research Group, Department of Mechanical Engineering, Faculty of Engineering and Information Technology, The University of Melbourne, Melbourne, Australia.
}%

\author{Ronald G. Larson}
\affiliation{%
Department of Chemical Engineering, University of Michigan, Ann Arbor, Michigan, USA.
}%

\author{Elnaz Hajizadeh}
\altaffiliation[]{Author to whom correspondence should be addressed; electronic mail: ellie.hajizadeh@unimelb.edu.au.}
\affiliation{ 
Soft Matter Informatics Research Group, Department of Mechanical Engineering, Faculty of Engineering and Information Technology, The University of Melbourne, Melbourne, Australia.
}%

\date{\today}

\begin{abstract}
Predicting the rheological behavior of associative polymers bridging colloidal particles into transient networks remains a fundamental challenge because the vast range of coupled spatiotemporal scales means no existing simulation technique can efficiently model these systems while accounting for critical molecular fidelity. We address this through development of a novel unified multiscale simulation framework for telechelic hydrophobically modified ethoxylated urethane (HEUR) polymer-colloid suspensions integrating: explicit-chain Brownian dynamics simulations that resolve polymer-particle association kinetics, active learning metamodels compressing these kinetics into computationally efficient surrogates across formulation space, and Population Balance-Brownian Dynamics (Pop-BD) simulations ingesting metamodel predictions to compute network-scale dynamics. Validated against explicit-chain Brownian dynamics, our framework accurately reproduces time- and frequency-dependent stress relaxation moduli, while enabling the simulation of larger systems over longer timescales. The framework achieves an 11-fold reduction in computational time compared to explicit-chain simulations while using significantly less expensive CPU resources instead of GPU hardware. Systematic investigations reveal that network connectivity exhibits critical transitions at specific chain-to-particle ratios, with bond (bridge) density and life-time directly correlating to enhanced relaxation times and moduli. Bond autocorrelation analysis also reveals that higher particle volume fractions yield more persistent bonds and slower relaxation. The framework successfully connects chain-level association dynamics to macroscopic rheology, providing a computationally efficient and physically informed approach for the rational design of associative colloidal materials, with direct applications to waterborne coating formulations and broader soft matter systems.

\end{abstract}
\maketitle

\section{Introduction}\label{sec:level1}

Associative polymer-colloid suspensions, in which associative polymers interact with colloidal particles to form dynamic networks, constitute a class of multi-scale materials with widespread importance across biological and industrial applications. These systems play essential roles in diverse contexts, including biological systems \cite{chen2013structure}, drug delivery \cite{chen2013structure}, food product formulations \cite{kreuter2014colloidal}, personal care products \cite{xu2019polymer}, and waterborne coatings \cite{ginzburg2018oscillatory}. The latter application, waterborne coatings, has 
been in demand over the past half-century, serving as a more sustainable alternative to conventional coatings where toxic volatile organic compounds (VOCs) are commonly present \cite {european2004directive}.


Waterborne coatings are characterized by water as their solvent, instead of VOCs, and primarily consist of distinctive combinations of colloidal latex particles, pigments, surfactants, and polymeric rheology modifiers (RM) in varying concentrations \cite{van2015shear}. Among them, the interaction between colloidal latex particles and polymeric rheology modifiers has been proven to be of significant importance when studying the rheology of Waterborne coatings. Latex particles are spherical polymer particles that range from approximately 50 to 300 nm in diameter. These particles constitute at least half of the weight of the waterborne coating formulation and significantly influence its mechanical properties\cite{de2010high}. RMs, however high-cost, also play a crucial role in controlling viscosity and shaping a desired shear-dependent viscosity. They fall into two categories: simple thickeners, like hydroxyethylcellulose (HEC), which increase viscosity by occupying space \cite{kastner2001impact}, and associative thickeners (ATs), such as hydrophobically modified ethylene oxide urethanes (HEUR). ATs are capable of associating and interacting with hydrophobic surfaces such as themselves and colloidal particles, where their association with pigment particles, such as TiO$_2$, is considered negligible \cite{van2013shear}. HEURs are telechelic polymers that feature a long hydrophilic poly(ethylene oxide) (PEO) middle block capped with hydrophobic alkyl groups, a.k.a. as "stickers" at each end. These stickers primarily associate with (stick to) hydrophobic surfaces, i.e., themselves and colloidal particles.  

In the absence of colloidal particles and at concentrations greater than critical micelle concentration ($C>C^*$), HEUR polymers self-assemble into flower-like micelles comprising hydrophobic cores enveloped by extended hydrophilic coronae.  At moderate concentrations, stickers entering and leaving the adjacent flowers enable network relaxation. The most straightforward theory for such solutions was developed by Green and Tobolsky \cite{green1946new} and termed the “temporary network model”. In this relatively simple theory, the network relaxation is determined by a single relaxation time, denoted as $\tau$. When there is no flow, this relaxation time is exponentially dependent on the free energy cost, $\Delta G$, for the escape of the sticker from the micelle, given by the equation $\tau = \tau_0 \exp (\frac{\Delta G}{k_B T})$.  According to the work of Tanaka and Edwards \cite{tanaka1992viscoelastic}, $ \tau_0=\omega_0^{-1} \approx$ 1-10 $ns$, where $\omega_0$ is a characteristic vibration frequency which refers to a rate at which a sticker attempts to overcome the potential barrier. $\Delta G $ was found by Annable \textit{et al.} \cite{annable1993rheology} to be linearly increasing to the strength of the sticker length ($C_n$) with a rate of approximately 1$ k_B T$  per methylene group at room temperature. As a result, the viscoelastic response of a typical HEUR to an applied oscillatory deformation follows a Maxwell model with a single relaxation time.

The rheological response of HEUR-thickened latex dispersions exhibits markedly different characteristics from pure polymer solutions, with colloid addition primarily influencing low-frequency storage modulus behavior while leaving high-frequency dynamics relatively unchanged \cite{pham1999polymeric}. At dilute particle loadings, modest deviations in $G\;' $ at low frequencies indicate weak particle clustering without extensive network formation, whereas increasing volume fraction enhances $G\;' $ sensitivity to colloid concentration and can drive crossover between storage and loss moduli \cite{chatterjee2017formulation}, signaling gel-like behavior with multiple characteristic relaxation timescales spanning colloidal cluster rearrangement ($\omega \approx$ 0.1–1 rad/s) to polymer bridge-loop transitions ($\omega \approx$ 10–100 rad/s) \cite{rubinstein2003polymer}. This broad relaxation spectrum and deviation from terminal scaling ($G\;' \sim \omega^2$, $G\;''  \sim \omega$) demonstrates that simple single-mode Maxwell models inadequately capture the complex dynamics of these polymer-bridged colloidal networks \cite{chatterjee2014shear}.  This complex rheological behaviour of latex-HEUR suspensions motivated computational  studies to uncover polymer-colloid organizations responsible for their structure-property relationships. Early theoretical frameworks proposed that HEUR polymers and latex particles form coexisting networks, wherein hydrophobic stickers partition between micellar junctions and particle surfaces, incorporating latex as additional network nodes alongside flower-like micelles \cite{pham1999micellar}.  However, Beshah \textit{et al.} \cite{beshah2013diffusion} used diffusion-weighted pulsed field gradient NMR to conclude that the vast majority of stickers are absorbed onto colloid surfaces at typical industrial paint formulation concentrations, indicating that the HEUR chains would primarily be in loop or bridge configurations on colloidal particles. This finding has been also confirmed by simulation studies \cite{ginzburg2015modeling}.  

Simulating these systems is challenging due to the vast differences in length and time scales among its constituents. Polymer dynamics processes, such as chain conformational changes and steric interactions, occur on timescales ranging from nanoseconds to seconds, while macroscopic rheological behavior arises from the chain-mediated collective network rearrangements of many particles over micron lengths and seconds. No single simulation \cite{MD4,MD5,MD6,Amal1,Amal2,Amal3} method can efficiently capture this complete spectrum of dynamics. Explicit Brownian dynamics (BD) simulations \cite{sg1,sg2,sg3} that resolve individual polymer chains at the Kuhn segment level provide accurate microscopic detail but become computationally prohibitive for systems containing more than a few dozen particles or simulation times exceeding a few seconds \cite{rezvantalab2018bridging,travitz2021brownian}. In intermediate coarse-graining, colloid/dumbbell simulations model polymer chains as finitely extensible nonlinear elastic springs (dumbbells) that connect adsorbing beads. This approach reduces the degrees of freedom sufficiently to simulate small multi-particle systems and extract stress relaxation moduli through time-autocorrelation analysis \cite{wang2018multiple}. These simulations identified four distinct relaxation modes, including polymer chain relaxation ($\tau_{HEUR} $), surface diffusion ($\tau_{surf}$), bridge dissociation ($\tau_{bridge} $), and cluster rearrangement ($\tau_{cl} $) and demonstrated shear-rate-dependent viscosity and bridge populations under flow \cite{krishnamurthy2023brownian}. To address the computational challenges posed by the numerous degrees of freedom, Hajizadeh et al. developed a novel mesoscopic hybrid method called population balance-Brownian dynamics (Pop-BD). This approach replaces the explicit tracking of polymers with probabilistic population balance equations that evolve the numbers of loops and bridges based on their transition rates, enabling the simulation of large systems over time scales relevant to experimental conditions. Although it sacrifices some resolution of rapid internal chain dynamics, this approach offers a more efficient method for simulating large systems. The recent integration of Pop-BD with HOOMD-blue software, incorporating efficient on-the-fly autocorrelation algorithms \cite{travitz2021multiscale}, has further enhanced scalability.

The predictive capability of Pop-BD depends critically on accurate transition rates between loop and bridge configurations, which govern the probabilistic evolution of polymer-colloid network topology and ultimately determine rheological response. Previous implementations of Pop-BD \cite{hajizadeh2018novel} have relied on simplified analytical expressions for transition rates\cite{tripathi2006rheology}, assuming transition frequencies scale exponentially with sticker binding energy and chain stretching. While these expressions capture qualitative trends, they neglect essential physics including particle curvature effects, finite polymer extensibility, many-chain interactions, and the complex multidimensional dependence on system parameters. Obtaining accurate transition rates requires fine-grained BD simulations that explicitly resolve polymer chains at the Kuhn segment level. 

These transition rates depend on a range of formulation variables, including chain length ($N_k$), particle radius ($R_p$), sticker binding energy ($\varepsilon_s$), polymer density ($n_{pol}/n_{col}$), and particle volume fraction ($\phi$), representing the gap between particles. Developing a quantitative understanding of these dependencies through conventional full factorial design would require $O(10^3)$ individual simulations.  Since each fine-grained BD simulation is computationally expensive (requiring days), this combinatorial approach would demand prohibitive computational resources, motivating our earlier work \cite{abdolahi2025interpretable} to develop an efficient and intelligent strategy for exploring the vast and multi-dimensional parameter space.

To replace conventional resource-intensive and exhaustive search strategy, in our previous study \cite{abdolahi2025interpretable}, we developed a computational framework integrating fine-grained Brownian dynamics simulations with active learning metamodeling. The BD simulations explicitly resolve polymer conformations at the Kuhn segment level and eliminate common simplifying assumptions such as flat-surface geometries \cite{travitz2021brownian}, enabling accurate prediction of transition rates across the full parameter space. 
We utilized these explicit BD simulations in an active learning metamodelling strategy to to develop a global surrogate model mapping those formulation variables and inter-colloid distance to transition rates between loop and bridge states. The active learning strategy allowed us to intelligently sample the space of input variables to minimize the number of expensive explicit BD simulations while maintaining global predictive accuracy for the transition rates. The resulting metamodels can provide Pop-BD with physically grounded, data-driven transition rates that capture the complete dependence on chain length, particle size, inter-particle spacing, polymer surface density, and sticker hydrophobicity, enabling accurate predictions of rheological behavior at larger time and length scales.

\begin{figure*}
\includegraphics[width=0.65\textwidth]{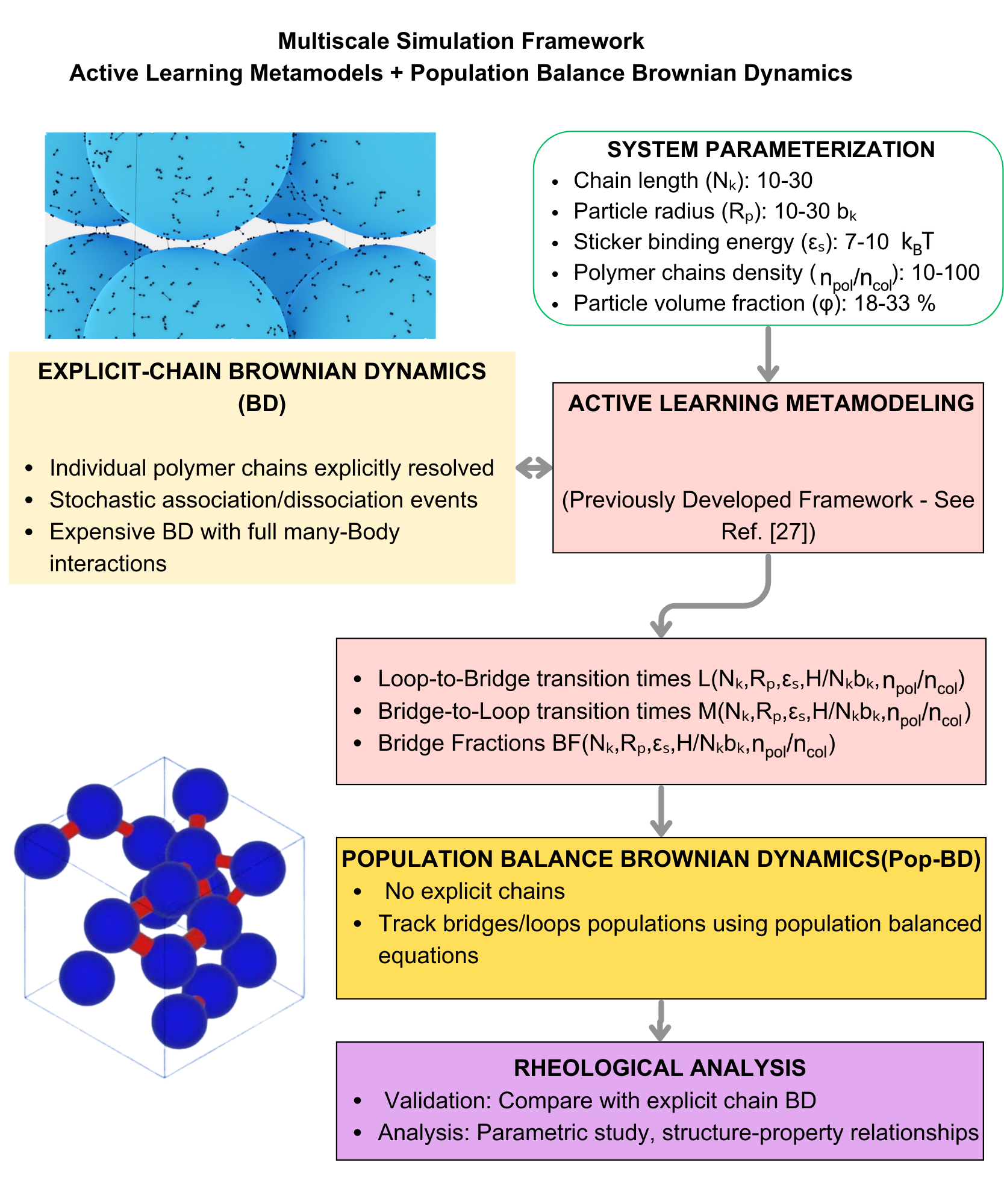}
\caption{\label{fr0}Novel ML-enabled multiscale simulation framework integrating fine-grained explicit chain Brownain dynamics (BD) simulations with Population Balance-Brownian dynamics (Pop-BD) through active learning metamodeling. The workflow begins with system parametrization defining the five-dimensional parameter space (chain length $N_k$, particle radius $R_p$, sticker binding energy $\varepsilon _s$, polymer surface density $n_{pol}/n_{col}$, and particle volume fraction $\phi$, which can be converted to normalized gap between particles, $H/N_kb_k$, in explicit chain BD simulations). Explicit-chain Brownian dynamics simulations resolve individual polymer chains and capture stochastic association/dissociation events, providing transition times and bridge fractions. These data inform Gaussian process-based active learning metamodels (previously developed, Ref. \cite{abdolahi2025interpretable}) that efficiently and intelligently map the parameter space with an order of magnitude less computation compared to conventional exhaustive search methods. The resulting metamodels serve as input for coarse-grained Pop-BD simulations, which track bridge/loop populations without explicitly resolving polymer chains, enabling access to larger timescales. Rheological analysis validates predictions against explicit-chain simulations and explores structure-property relationships. Left top: Schematic visualization of the polymer-bridged colloidal network structure in explicit-chain BD simulations. Left bottom: Pop-BD representation where explicit chains are replaced with dynamical bonds, enabling simulation of larger systems and timescales.}
\end{figure*}

In this work, we present a comprehensive multiscale simulation framework (Figure \ref{fr0}) that integrates active learning metamodels with Population Balance-Brownian Dynamics simulations to predict the rheological behavior of associative polymer-colloid suspensions and identify the underlying polymer-particle associations driving them. We employed the latest implementation of the Pop-BD method within the high-performance HOOMD-blue simulation platform with  on-the-fly autocorrelation algorithms that eliminate the need for trajectory storage during stress relaxation calculations \cite{travitz2021multiscale}.  

We make three principal contributions that collectively enable predictive simulation of network dynamics across previously inaccessible spatiotemporal scales. First, we integrate multi-variable transition rates and equilibrium bridge fractions derived from metamodels trained through active learning into the Pop-BD framework, replacing the simplified analytical expressions based on entropic spring theory with physically accurate, data-driven transition rates extracted from fine-grained BD simulations. Second, we validate this ML-enhanced Pop-BD approach by comparing computed linear stress relaxation moduli against explicit-chain BD simulations for benchmark systems, demonstrating that Pop-BD accurately reproduces long-time network relaxation dynamics while remaining computationally tractable for large particle numbers and extended simulation times. Third, we exploit the computational efficiency of ML-enhanced Pop-BD to investigate systems containing tens to hundreds of particles over timescales spanning multiple network relaxation events, systematically examining how formulation parameters, including colloid volume fraction, polymer chain length, and sticker attraction strength, influence network connectivity, bridge population dynamics, and rheological properties. These contributions establish a validated multiscale framework that bridges molecular-level association kinetics with system-level rheological performance, providing a practical tool for computational materials design of waterborne coating formulations.

The remainder of this paper is organized as follows: Section \ref{back} describes the computational methodology, including the Brownian dynamics simulation model and Pop-BD implementation. After discussing small-scale relaxation times through explicit chain Brownian Dynamics (BD) simulations in Subsection \ref{R1}, Subsection \ref{R2} presents validation results that compare Pop-BD with the explicit chain simulations. This subsection also demonstrates how the framework can be applied to larger systems across a range of formulation conditions.  Subsection \ref{R3} discusses the physical insights gained regarding network formation, relaxation mechanisms, and rheological behavior. Section \ref{con} concludes with implications for coating formulation and future research directions.

\section{Theoretical Background}\label{back}
Figure \ref{fr0} illustrates our novel machine learning-enabled multiscale simulation framework, which seamlessly integrates explicit-chain Brownian dynamics, active learning
metamodeling, and population balance equations to predict rheological properties across multiple length and time scales. In what follows, we will discuss the individual components of this integrated multiscale framework in great detail. 

\subsubsection{Explicit chain Brownian Dynamics Model}

In our previous work \cite{abdolahi2025interpretable}, we developed a detailed Brownian Dynamics (BD) simulation framework to study the association dynamics between hydrophobically ethoxylated urethane (HEUR) polymers and colloidal latex particles at the Kuhn step resolution, enabling accurate predictions of transition rates across the full range of particle gap-to-radius ratios. The simulated system features a ``minimal model'' consisting of a periodic simulation box containing polymeric chains and a fixed particle positioned at the center. This represents the smallest unit capable of simulating bridge and loop transition dynamics, with periodic images of the particle serving as neighboring particles for bridge formation.  Colloidal particles with hard-core radius $R_p$ ranging from 11 to 33 nm, and HEUR polymeric chains modeled as bead-spring chains. The poly(ethylene oxide) (PEO) backbones of the HEURs is represented as a freely-jointed Kuhn chain with segment molecular weight of 137 g/mol and effective length ($b_k$) of approximately 1.1 nm \cite{chatterjee2014shear}. The number of Kuhn steps ($N_k$) was varied from 10 to 30, corresponding to end-to-end distance to colloid radius ratios between 0.1 and 0.5, which aligns with experimental ratios ($\sim$0.23) \cite{ginzburg2018oscillatory}. The number of chains per particle ranged from 10 to 100, approaching typical experimental values \cite{chatterjee2017formulation}.

In such a model, the bond potential between beads is described as:
\begin{equation}\label{eq_utility_401}
U_{\text{bond}}(r) =\frac{1}{2}k(r-r_0)^2,
\end{equation}
where $k = 400~k_BT/b^2_k$ and $r_0 = 1.0~b_k$.  A Weeks-Chandler-Anderson (WCA) potential, 

\begin{equation}\label{eq:bead_bead}
U_{\text{bb}}(r) =
\begin{cases} 
 4\varepsilon_{\text{bb}}\left[\left(\frac{\sigma_{\text{bb}}}{r}\right)^{12} - \left(\frac{\sigma_{\text{bb}}}{r}\right)^{6} + \frac{1}{4}\right], & r < 2^{1/6}\sigma_{\text{bb}} \\  
 0, & r \geq 2^{1/6}\sigma_{\text{bb}}
\end{cases}
\end{equation}
where $\varepsilon_{\text{bb}} = 1.0~k_BT$ and $\sigma_{\text{bb}}= 0.4~b_k$, describes the bead-bead interaction potential. The stickers can adsorb on the colloidal particles according to a shifted Lennard-Jones potential,
\begin{equation}\label{eq:sticker_particle}
U_{\text{ps}}(r) = 
\begin{cases} 
4\varepsilon_{s}\left[\left(\frac{\sigma_{\text{ps}}}{r - R_p}\right)^{12} - \left(\frac{\sigma_{\text{ps}}}{r - R_p}\right)^{6}\right], & r < R_p + 2.5\sigma_{\text{ps}} \\  
0, & r \geq R_p + 2.5\sigma_{\text{ps}}
\end{cases}
\end{equation}
where $\sigma_{\text{ps}} = 1.0~b_k$, $\varepsilon_s = 7$--$10~k_BT$, and $R_p$ is the hard-core particle radius. However, the non-sticker bead-particle interaction is modeled using a WCA potential, 
\begin{equation}\label{eq:bead_particle}
U_{\text{pb}}(r)= 
\begin{cases} 
 4\varepsilon_{\text{pb}}\left[\left(\frac{\sigma_{\text{pb}}}{r-R_p}\right)^{12}-\left(\frac{\sigma_{\text{pb}}}{r-R_p}\right)^{6}+\frac{1}{4}\right], & r< R_p+2^{1/6}\sigma_{\text{pb}}\\  
 0, & r\geq R_p+2^{1/6}\sigma_{\text{pb}}
\end{cases}
\end{equation}
where $\sigma_{\text{pb}} = 1.0~b_k$ and $\varepsilon_{\text{pb}} =1.0~k_BT$.

While these explicit-chain BD simulations provide detailed insight into polymer-colloid association dynamics and accurately capture transition kinetics, they remain computationally expensive for systems with hundreds of particles and chains per particle typical of experimental suspensions. This limitation motivated the development of the active learning metamodelling framework, employing Gaussian Process (GP) regression combined with adaptive sampling (a.k.a., Bayesian approach) to efficiently map how loop-to-bridge and bridge-to-loop transition rates depend on key system parameters, while reducing the required number of simulations from thousands to hundreds. These metamodels provide the transition rate data needed to inform the Pop-BD framework, which replaces explicit polymer chains, and will be discussed in the next section, allowing us to simulate large-scale systems with hundreds of particles and chains per particle, approaching experimentally relevant timescales while preserving the essential physics of polymer-particle interactions. 

\subsubsection{Population Balance Brownian Dynamics Method}\label{back2}

Population balance Brownian dynamics (Pop-BD) \cite{hajizadeh2018novel} is a hybrid mesoscopic simulation approach that combines population balance equations for the loop and bridge configurations with Brownian dynamics simulation of colloidal particles to examine the behaviour of complex suspensions containing colloidal particles and associative polymers. This method is particularly effective for modelling polymer-bridged colloidal latex suspensions, where simulating every polymer molecule explicitly would be computationally prohibitive. In these systems, telechelic polymers function as reversible linkers between particles, forming a transient network that influences the phase behaviour and rheological properties of the suspension. Unlike fully explicit Brownian Dynamics simulations, which track both colloidal particles and polymer chains individually, the Pop-BD method offers a significant advantage by focusing on the colloidal particles and the number of polymer bridges connecting them, rather than tracking each individual polymer chain. 

In Pop-BD, polymer chains are implicitly represented by a \(N_{\text{particle}} \times N_{\text{particle}}\) matrix. The off-diagonal elements of this matrix denote the number of bridges \(N_{ij}\) connecting particles \(i\) and \(j\), while the diagonal elements indicate the number of loops associated with each particle. Population balance equations are then employed to compute the number of bridges \(N_{ij}\) between the \(i^{\text{th}}\) and \(j^{\text{th}}\) particles, as well as the number of loops on the \(i^{\text{th}}\) and \(j^{\text{th}}\) particles (\(N_{ii}\) and \(N_{jj}\)) at any given time step, as formulated by:

\begin{equation}\label{eq_utility_56}
\frac{dN_{ij}}{dt} = (N_{ii} + N_{jj})  L(d_{ij}) - N_{ij} M(d_{ij}), 
\end{equation}
\begin{equation}\label{eq_utility_57}
\frac{dN_{ii}}{dt} = -N_{ii} L(d_{ij}) + \frac{N_{ij} M(d_{ij})}{2},
 \end{equation}
\begin{equation}\label{eq_utility_58}
 \frac{dN_{jj}}{dt} = -N_{jj}  L(d_{ij}) + \frac{N_{ij} M(d_{ij})}{2}.
 \end{equation}
In this context, $d_{ij}$ denotes the distance between the $i^{th}$ and $j^{th}$ colloids. The variables $M(d_{ij})$ and $L(d_{ij})$ represent the gap-dependent rates for transitioning from bridge to loop and from loop to bridge, respectively. 

The probability that a chain will transition from loop to bridge configuration after $\Delta t$ time step is $p=L(d_{ij}) \Delta t$, and that of bridge-to-loop transition is $q=M(d_{ij}) \Delta t$. Thus, the probability of forming a single bridge between two neighbouring particles \(i\) and \(j\) due to the de-adsorption of a loop, \(P_{ij}\), and that of forming a loop as a result of the breakage of  a bridge, $Q_{ij}$ are:
\begin{equation}\label{eq_utility_511}
P_{ij} = N_{ii} p \left(1 - p \right)^{N_{ii}  - 1},
\end{equation}

\begin{equation}\label{eq_utility_512}
Q_{ij} = N_{ij} q \left(1 - q \right)^{N_{ij} - 1}.
\end{equation}

With the probabilities \(P_{ij}\), \(P_{ji}\), and \(Q_{ij}\) calculated, the following algorithm is used to update the number of loops and bridges in a single discrete time-step,  as: 

\begin{itemize}
    \item Compute \(P_{ij}\), \(P_{ji}\), and \(Q_{ij}\).
    \item Generate four uniformly distributed random numbers, \(l_1\), \(l_2\), \(l_3\), and \(l_4\), each in the range \([0, 1]\).
    \item If \(l_1 < P_{ij}\), then:
    \begin{itemize}
        \item Update \(N_{ii} \gets N_{ii} - 1\)
        \item Update \(N_{ij} \gets N_{ij} + 1\)
    \end{itemize}
    \item If \(l_2 < P_{ji}\), then:
    \begin{itemize}
        \item Update \(N_{jj} \gets N_{jj} - 1\)
        \item Update \(N_{ij} \gets N_{ij} + 1\)
    \end{itemize}
    \item If \(l_3 < Q_{ij}\), then:
    \begin{itemize}
        \item If \(l_4 \leq 0.5\), then:
        \begin{itemize}
            \item Update \(N_{ii} \gets N_{ii} + 1\)
            \item Update \(N_{ij} \gets N_{ij} - 1\)
        \end{itemize}
        \item Else:
        \begin{itemize}
            \item Update \(N_{jj} \gets N_{jj} + 1\)
            \item Update \(N_{ij} \gets N_{ij} - 1\)
        \end{itemize}
    \end{itemize}
    \item Repeat the steps for all particle pairs \(i < j\) in the system.
\end{itemize}
It is important to note that the Pop-BD algorithm updates one bridge per particle pair during each time step. However, the simulation time step, $\Delta t$, is intentionally kept low to satisfy $N_{ij} M(d_{ij}) \Delta t \ll 1$, ensuring that the probability of multiple bridge formation or breakage events occurring between the same particle pair within a single time step is negligible. Additionally, $\Delta t$ is at least three orders of magnitude smaller than the characteristic diffusion time needed for a particle to move a distance comparable to the interaction range, which prevents particle overlap. Following the calculation of the number of bridges at each time step, equivalent spring forces \( U_s \) are assigned between pairs of colloidal particles, utilizing the Cohen--Pad\'e approximation \cite{cohen1991pade}, along with the purely repulsive Weeks-Chandler-Andersen potentials, \( U_{pp} \), as follows:

\begin{equation}\label{eq_utility_us}
U_s(d) = N_K k_B T \left[ \frac{1}{2} \left( \frac{d}{N_K \times b_K} \right)^2 - \ln \left( 1 - \left( \frac{d}{N_K \times b_K} \right)^2 \right) \right],
\end{equation}

\begin{equation}\label{eq_utility_514}
    U_\text{pp}(r) = 
    \begin{cases} 
      4\varepsilon_\text{pp} \left[ \left( \frac{\sigma_\text{pp}}{r - D} \right)^{12} - \left( \frac{\sigma_\text{pp}}{r - D} \right)^6 \right] + \frac{1}{4}, & r < D + r_D, \\
      0, & r \geq D + r_D.
    \end{cases}
\end{equation}
In this context, \(d\) denotes the elongation of the polymer chain or spring, indicating how far the chain is stretched. $N_K \times b_K$ signifies the maximum extension of the spring, where \(b_K\) represents the length of a Kuhn segment and \(N_K\) indicates the quantity of Kuhn steps. \(r\) refers to the distance from one colloidal particle's centre to another's centre. \(D\) is the diameter of the particles, establishing the closest distance at which the particles start to interact. \(r_D\) is the cut-off distance for the potential, ensuring that interactions take place only up to \(D + r_D\). \(\varepsilon_{pp}\) indicates the potential well depth, which reflects the intensity of the repulsive interaction. \(\sigma_{pp}\) is the parameter for the range of the Lennard-Jones potential and set to 2.0 \(b_K\) \cite{wang2018multiple}, which adjusts the distance component in the potential.

 Pop-BD simulations were performed using a custom extension to HOOMD-blue v2.9 \cite{anderson2020hoomd}, incorporating runtime bond modification and on-the-fly stress autocorrelation \cite{travitz2021multiscale}. This approach eliminates trajectory storage requirements and enables efficient calculation of the linear relaxation modulus, $G(t)$, for long simulations. In the linear response regime, $G(t)$ is defined as the ratio of the stress response to an applied small strain, under the assumption that the response is directly proportional to the perturbation. For isotropic systems, $G(t)$ is often expressed through the stress-stress autocorrelation function:
\begin{equation}\label{eq_utility_2}
G(t)=\frac{V}{k_BT}<S_{xy} (t)S_{xy} (0)>,
\end{equation}
where V is the volume of a periodic box, and $S_{xy}(t)$ is the shear stress in $ xy$ plane at time $t$, $S_{xy}(0)$  is the $xy$ shear stress at a reference time 0, and $<>$ denotes ensemble averaging over the simulation duration after equilibration. The stress component, $S_{xy}(t)$, is calculated by forces between dynamical bridges (implicit chains in Pop-BD stimulations), while ignoring the force contributions from particle-particle interactions  to minimize the noise in $G(t)$, aligned with Wang and Larson \cite{wang2018multiple} studies.





\section{Results and Discussions} \label{Results}
\subsubsection{Small-Scale relaxation times through explicit chain BD simulations}\label{R1}
\begin{figure}
    \centering
    \includegraphics[width=0.4\textwidth]{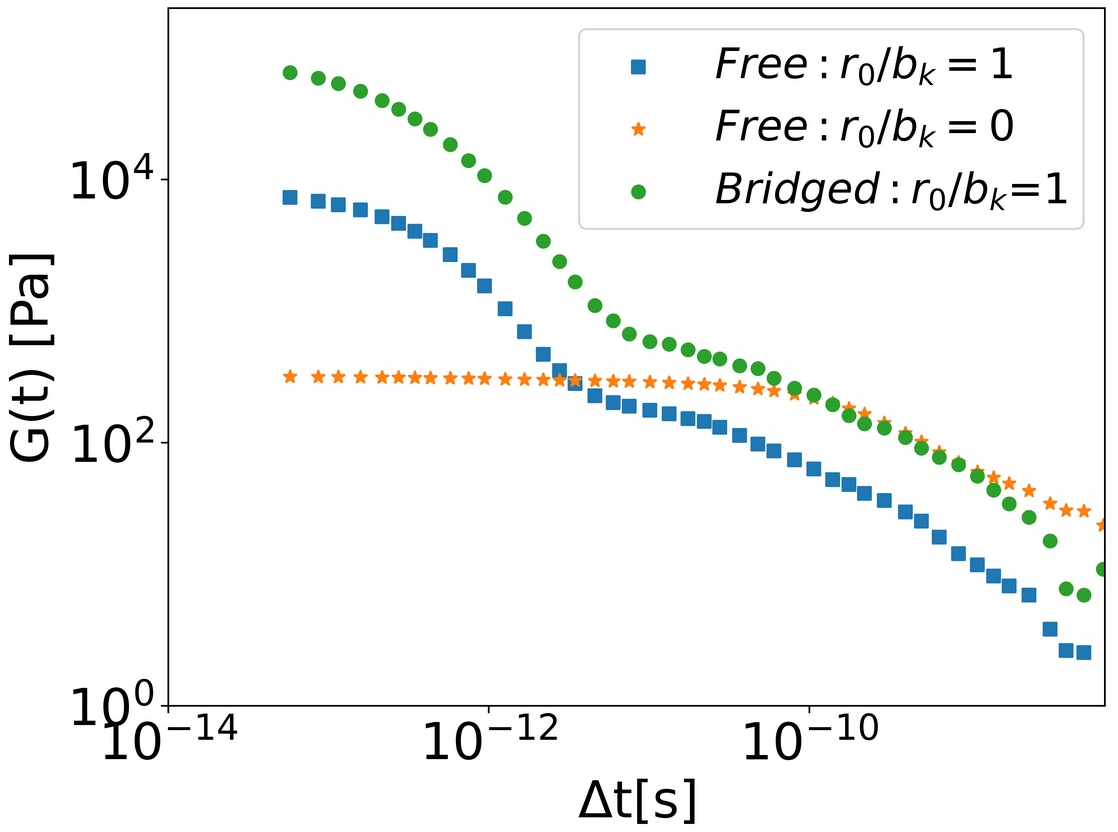}
    \caption{\label{f601} The stress relaxation modulus, \(G(t)\), for a single polymer chain with a non-zero rest length \(r_0\) (represented by blue squares), a zero rest length (\(r_0 = 0\) and $k = 3k_BT/b_k^2$, shown by orange stars), and a bridged configuration is analysed. The stress relaxation modulus \(G(t)\) for a single chain with a resting length of \(r_0 = 1 b_k\) and $k = 400k_BT/b_k^2$ demonstrates a double decay pattern. This pattern is attributed to rapid bond-level relaxation followed by slower chain conformational changes. In contrast, when the resting length \(r_0\) is zero, \(G(t)\) exhibits a single slower decay, which is due to chain reconfiguration dynamics. The bridged chain with \(r_0 = 1 b_k\) and $k = 400k_BT/b_k^2$ displays a stiffer modulus and a slightly longer relaxation time compared to the free chain. This behaviour results from the restricted motion of the chain and its enhanced load-bearing capability.}
\end{figure}

In the first scenario, a single chain of \(N_k = 20\) (or 21 beads) is modeled in a box of \(60 \, b_k \times 60 \, b_k \times 60 \, b_k\) with no particles present, where $b_k$ is the length of one Kuhn segment equal to 1.1 nm. The bonds between the beads are modelled by the harmonic spring model given in Equation \ref{eq_utility_401}, $U_{\text{bond}}(r) = \frac{1}{2}k(r - r_0)^2$. Two parameterizations are employed to isolate distinct relaxation mechanisms: for chains with nonzero rest length ($r_0 = 1.0b_k$), the spring constant is set to $k = 400k_BT/b_k^2$, creating stiff, rod-like bonds with small thermal fluctuations around the equilibrium length. For chains with zero rest length ($r_0 = 0$), the spring constant is set to $k = 3k_BT/b_k^2$ to ensure consistency with classical Rouse theory, where the  mean-square bond length $\langle r^2 \rangle = b_k^2$ correctly represents a freely fluctuating Gaussian segment. The stress relaxation modulus, $G(t)$, displays markedly different behavior depending on these parameters, as shown in Figure \ref{f601}. When $r_0 = 1.0 b_k$, $G(t)$ exhibits a double decay pattern. The first, rapid decay corresponds to bond-length relaxation where the nonzero rest length induces additional elastic stress that relaxes quickly through thermal fluctuations around $r_0$. This process is influenced by the bond length elastic constant \( k \). The second, slower decay is related to the relaxation of bond orientations and overall chain conformations. In contrast, when $r_0 = 0$ with $k = 3k_BT/b_k^2$, $G(t)$ exhibits a single, smooth decay consistent with Rouse theory, governed solely by chain reconfiguration dynamics. The absence of the fast bond-relaxation mode occurs because bonds have no displaced equilibrium position. 



Additionally, Figure \ref{f601} shows \(G(t)\) for a single chain with \(N_k = 20\), \(k = 400 \, k_BT/b_k^2\), and \(r_0 = 1.0 \, b_k\), which is bridged between two fixed particles that are spaced 10 \(b_k\) apart. The dimensions of the simulation box are \(70 \, b_k \times 70 \, b_k \times 70 \, b_k\). The initial configuration of the chain and the strength of the sticker attraction (\(\varepsilon_s = 20 \, k_BT\)) were intentionally selected to maintain the bridged configuration throughout the simulations. The bridged chain has two end-capping stickers that can diffuse on the opposing colloidal particles. As a result, \(G(t)\) exhibits greater values with a slightly longer relaxation time scale compared to a free single chain. This difference arises mainly from the restricted movement of the bridged chain due to the presence of colloidal particles, a phenomenon also noted in Ref. \cite{wang2018multiple}.

\begin{figure}

    \includegraphics[width=0.43\textwidth]{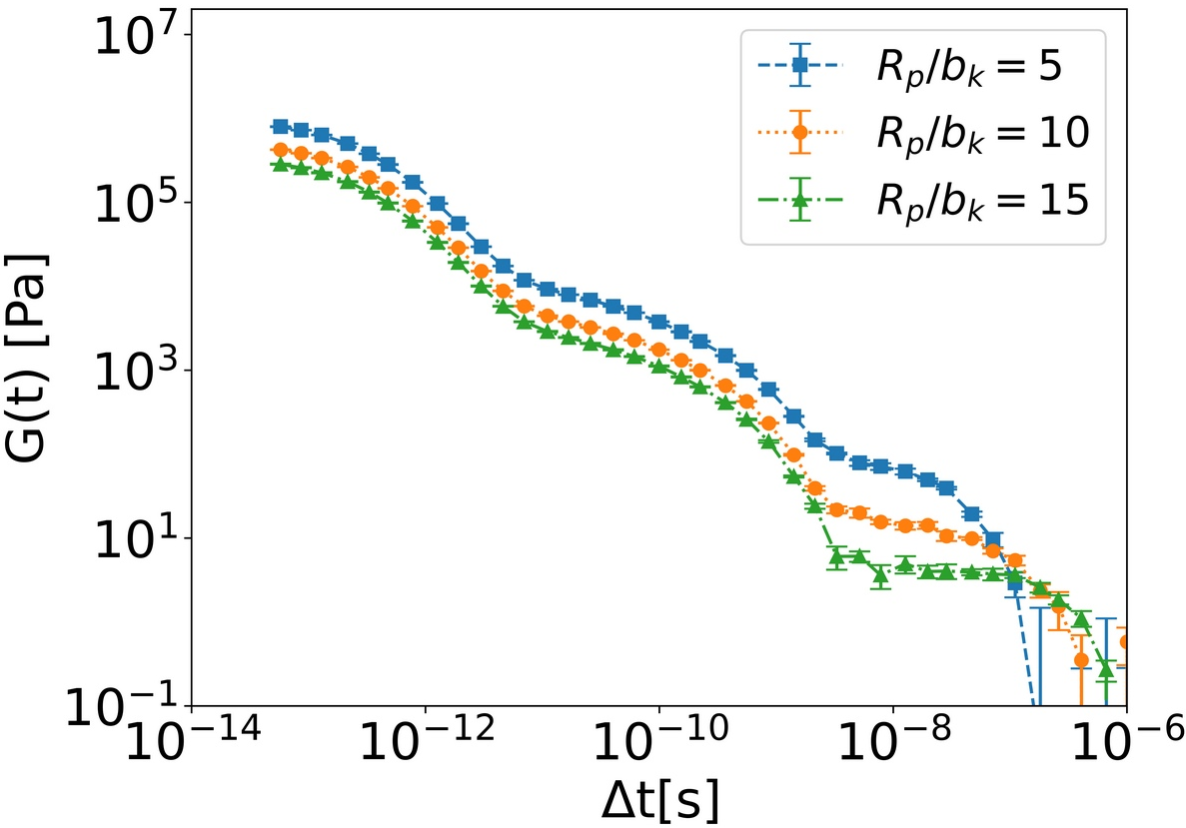}

   \caption{The stress relaxation modulus \( G(t) \) is analysed for systems with a fixed particle at the centre of a \( 60 \, b_k \times 60 \, b_k \times 60 \, b_k \) simulation box, containing a chain of \( N_k = 20 \). The chains have parameters \( k = 400 \, k_B T/b_k^2 \) and \( r_0 = 1.0 \, b_k \), with a sticker attraction strength of \( \varepsilon_s = 20 \, k_B T \). The longest relaxation time increases with particle radius, leading to longer loop relaxation times for larger particles. The other two relaxation times are the similar to those shown in Figure \ref{f601}. Each simulation was repeated five times with different random seeds, with error bars showing mean values and standard deviations.}
    \label{f602}
\end{figure}

Figure \ref{f602} displays the stress relaxation modulus for model systems that contain a fixed particle at the centre of a simulation box, with a dilute particle volume fraction of \(\phi = 5\%\) and a single chain in each box. Each chain consists of 10 Kuhn segments (\(N_k = 10\)), where the value of \(k\) is set to \(400 \, k_BT/b_k^2\), the bond length \(r_0\) is \(1.0 \, b_k\), and attraction strength \(\varepsilon_s\) of the sticker is set to \(20 \, k_BT\). The radius of the particles ranges from \(5\) to \(15 \, b_k\). Due to the dilute particle volume fraction and the high value of \(\varepsilon_s\), the stickers remain attached to the particle for the entire duration of the simulations. Initially, the chains are randomised and then equilibrated. The stress relaxation modulus \(G(t)\) for these systems includes a combination of relaxation times related to both chain relaxation and loop relaxation time, the latter of which describes how fast the stickers diffuse and slide across the surface of the particle. As illustrated in Figure \ref{f602}, the slowest relaxation time, associated with loop relaxation, is proportional to the radius of the particles, which means that an increase in the radius of the particle results in delayed loop relaxation times in agreement with previous findings in the literature\cite{wang2018multiple}.

\subsubsection{Validation: Active learning-informed Pop-BD vs. Explicit-chain BD}\label{R2}
\begin{figure}
    \centering
    \includegraphics[width=0.4\textwidth]{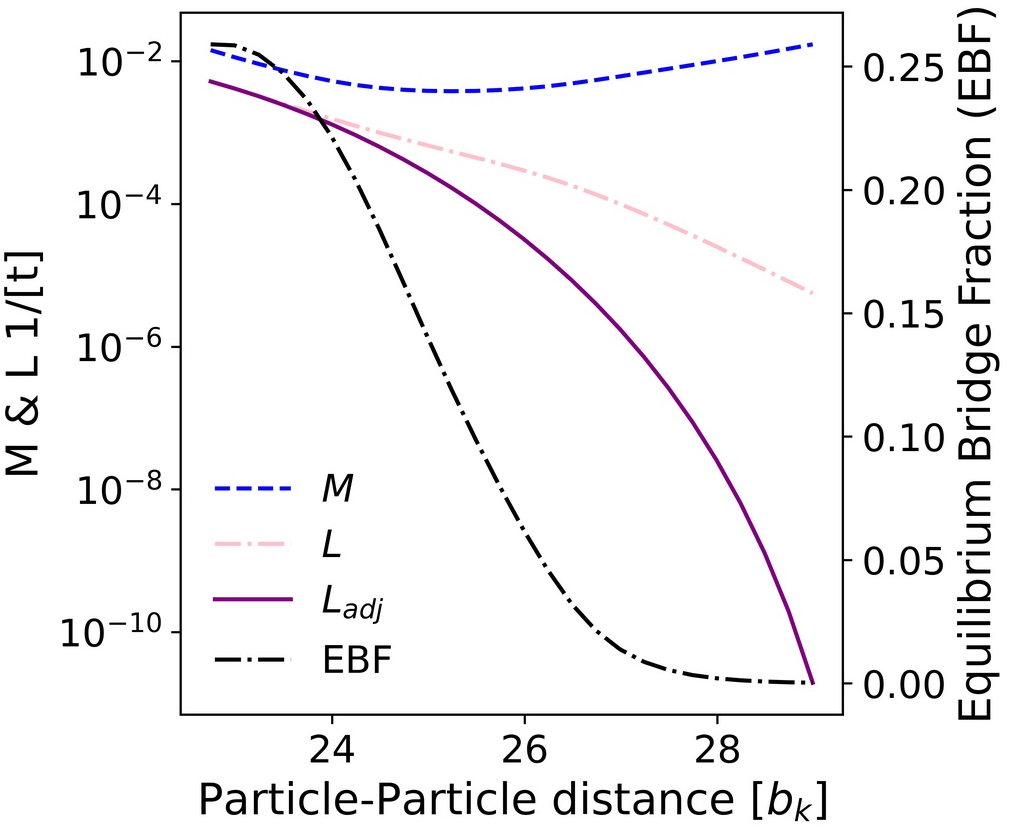}
    \caption{\label{f6000} Gap-dependent transition rates and equilibrium bridge fraction (EBF) for a system with $\varepsilon_s$ = 8 $k_B T$, $N_k = $10, $R_p/b_k $= 10, and $n_{pol}/n_{col}$ = 10. The bridge-to-loop transition rate $M$ (blue dashed line) and loop-to-bridge transition rate $L$ (pink dash-dot line) from active learning metamodels are compared with the adjusted loop-to-bridge rate $L_{adj}$ (purple solid line) that asymptotically matches the analytical expression (Equation \ref{eq_utility_601}) at large separations. The EBF (black dash-dot line, right axis) obtained from the metamodel decreases rapidly with increasing gap distance.}
\end{figure}

The inputs for Pop-BD simulations as discussed in Ref. \cite {travitz2021multiscale} are tabulated in the form of gap-dependent values for the loop-to-bridge transition rate, $L(d)$, the bridge-to-loop transition rate, $M(d)$, and the equilibrium bridge fraction, $EBF(d)$. Inspired by the Transient Network of Bridged Particles (TNBM) model\cite{ tripathi2006rheology} the original Pop-BD formulation\cite{hajizadeh2018novel} used below definitions for aforementioned quantities:
\begin{equation}\label{eq_utility_601}
L(d) = \Omega \exp \left( -\frac{1}{k_B T} \left[ \Delta G + U_s(d) \right] \right),
\end{equation}
\begin{equation}\label{eq_utility_602}
M(d) = \Omega \exp \left( -\frac{1}{k_B T} \left[ \Delta G \right] \right),
\end{equation}
\begin{equation}\label{eq_utility_603}
EBF(d) = \frac{L(d)}{L(d) + M(d)},
\end{equation}
where $d$ is the center-to-center interparticle distance, $\Omega$ is the thermal fluctuation frequency, $\Delta G \approx \varepsilon_s k_BT$ represents the association potential well depth, and $U_s(d)$ denotes the spring potential energy (Equation~\ref{eq_utility_us}).  In the present work, we extracted these quantities using the transition rates and equilibrium bridge fractions from surrogate ML models developed in our previous work from high-resolution explicit chain BD simulations \cite{abdolahi2025interpretable}. Figure \ref{f6000} shows an example of these previously developed metamodels for gap-dependent transition rates and equilibrium bridge fractions associated with a system with $\varepsilon_s = 8 \, k_B T$, $N_k = 10$, $R_p/b_k = 10$, and $N_c/N_p = 10$. As discussed in the work, the active learning acquisition function preferentially samples regions with higher bridge fractions; thus, the metamodel exhibits greater uncertainty at larger gaps where bridges are scarce. This is particularly evident in the loop-to-bridge rate $L$, which shows exponential sensitivity to particle separation (Figure \ref{f6000}). We therefore use an adjusted rate $L_{adj}$ that is equal to active learning predictions at small-to-moderate gaps while asymptotically matching Equation \ref {eq_utility_601} at large separations. In contrast, $M$ and $EBF$ are used directly from the metamodel without adjustment, as $M$ remains relatively constant and $EBF$ becomes negligible at large gaps.

\begin{figure}
    \centering
    \includegraphics[width=0.4\textwidth]{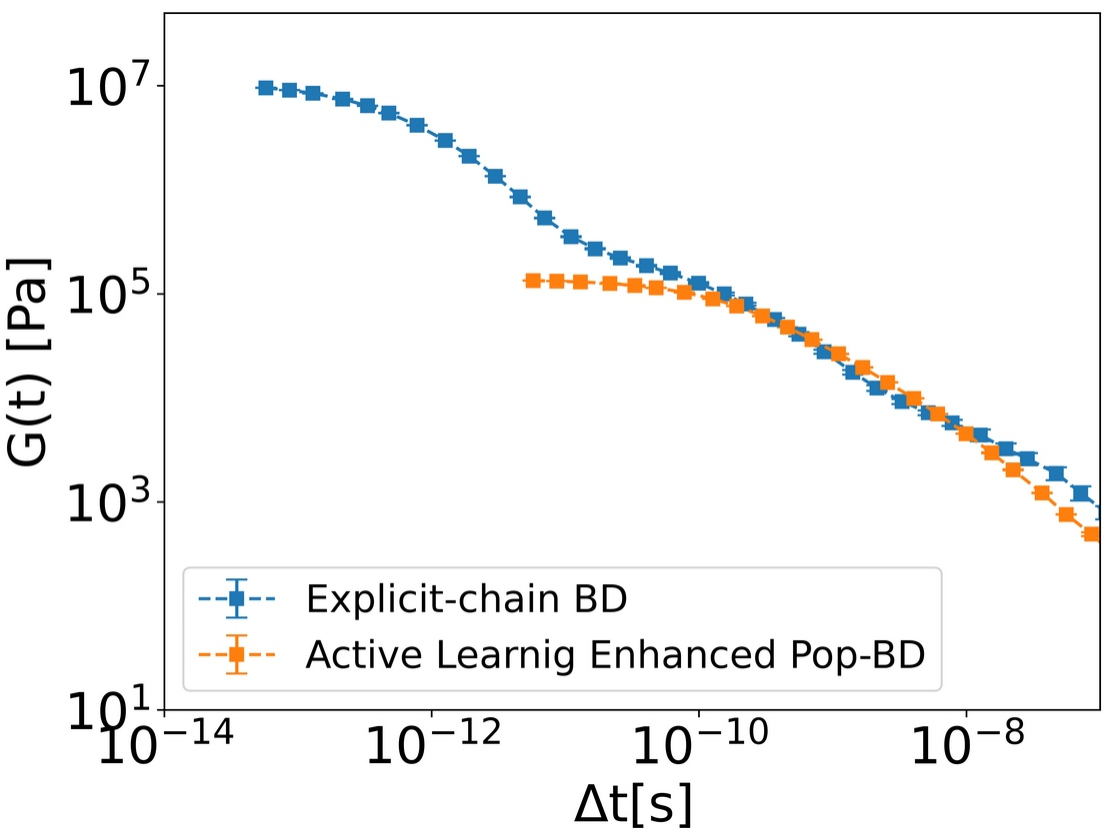}
    \caption{\label{fg1} Comparison of the stress relaxation modulus $ G(t)$ from Pop-BD simulations using active learning transition rates and bridge fractions shows excellent agreement with fully explicit-chain BD results. All simulations were conducted for a system of eight colloidal particles at a particle volume fraction of \(\phi = 20\%\), with \(\varepsilon_s = 8 k_B T\), \(N_k = 10\), \(R_p/b_k = 10\), and \(n_{pol}/n_{col} = 10\). Each simulation was repeated five times with different random seeds, with error bars showing mean values and standard deviations.}
    \end{figure}

The $L_{adj}$, $M$, and $EBF$ were used as inputs to Pop-BD simulations in this subsection with eight colloidal particles at a particle volume fraction of \(\phi = 20\%\). In these simulations, the implicit chains, with \(\varepsilon_s = 8 k_B T\), \(N_k = 10\), \(R_p/b_k = 10\), and \(n_{pol}/n_{col} = 10\),  were replaced by dynamic bonds between neighbouring particles, governed by the population balance equations (Equation \ref{eq_utility_56} -Equation \ref{eq_utility_512}). For comparison, we implemented a fully explicit-chain BD simulation for a similar system as the ground truth.  We compared the stress relaxation modulus, \(G(t)\), from the Pop-BD simulations integrated with active learning transition rates to that from the fully explicit-chain BD simulations, as illustrated in Figure \ref{fg1}.  The Pop-BD simulations, informed by active learning transition rates and bridge fractions, demonstrated excellent agreement with the fully explicit-chain BD results compared to the original versions of the transition rates. 
The Pop-BD method treats polymers implicitly through population balance equations rather than explicit chain coordinates. As expected, this coarse-graining \cite{Shireen2022,Weeratunge2023} eliminates short-time relaxation modes associated with polymer conformational dynamics, resulting in a \(G(t)\) curve that remains flat at early times compared to the rapid decay observed in fully explicit BD simulations.
To ensure statistical parity between Pop-BD and explicit BD simulations, each simulation was repeated five times using independently generated random seeds.

\begin{figure}
    \centering
    \includegraphics[width=0.4\textwidth]{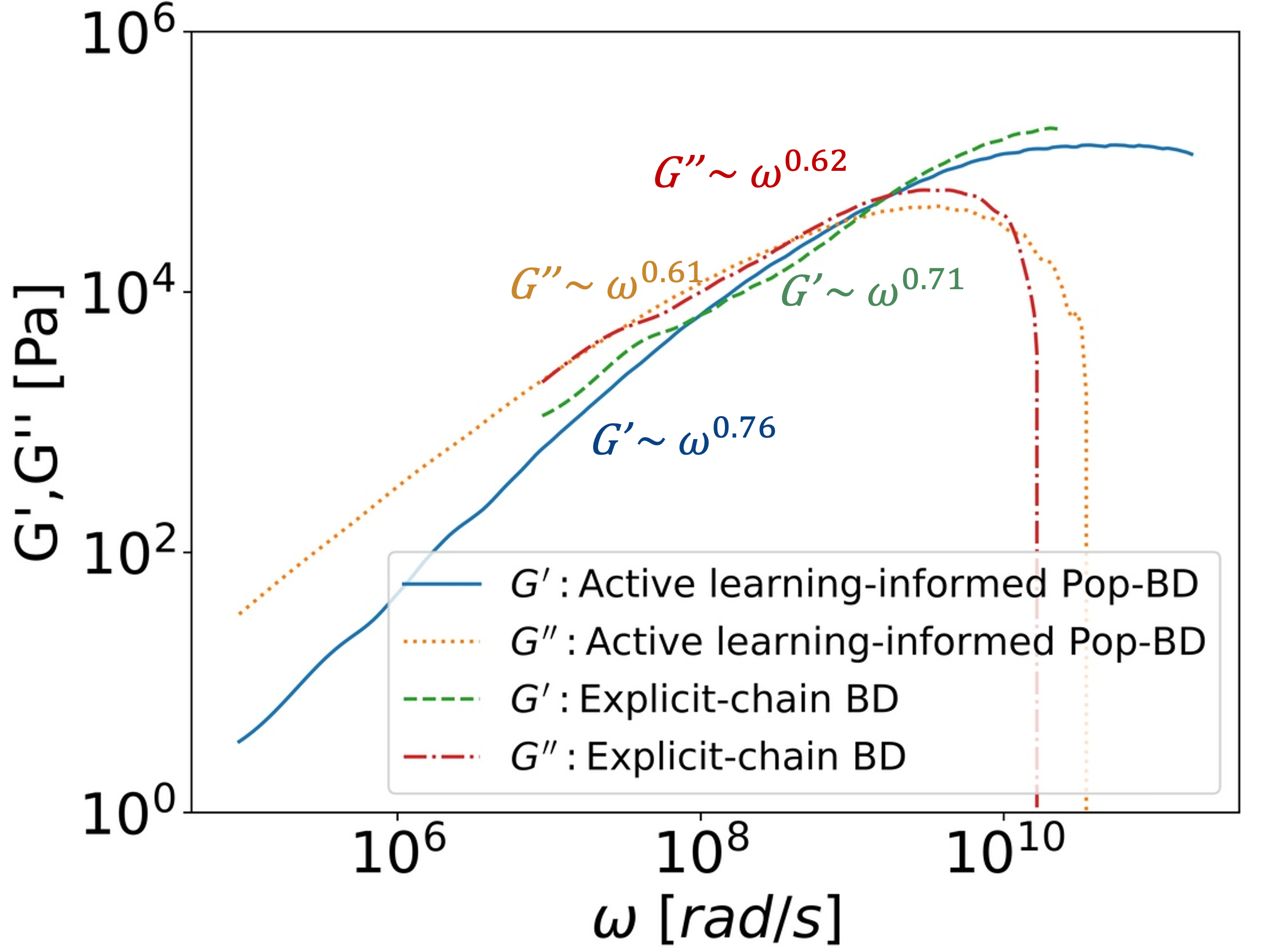}
    \caption{\label{fr1} Comparison of the stress relaxation modulus G(t) from Pop-BD simulations using active learning transition rates and bridge fractions shows excellent agreement with fully explicit-chain BD results. All simulations were conducted for a system of eight colloidal particles at a particle volume fraction of \(\phi = 20\%\), with \(\varepsilon_s = 8 k_B T\), \(N_k = 10\), \(R_p/b_k = 10\), and \(n_{pol}/n_{col} = 10\).}
    \end{figure}

To further validate the active learning-informed Pop-BD predictions, we converted the time-domain linear stress relaxation modulus $G(t)$ to frequency-dependent storage ($G\;'$) and loss ($G\;''$) moduli using the i-Rheo web application (\href{https://i-rheo.mib3avenger.com/}{https://i-rheo.mib3avenger.com/}) \cite{tassieri2018rheo}. This conversion implements an analytical method for evaluating Fourier transforms of sampled functions over finite time windows without requiring preconceived models such as generalized Maxwell fits. Specifically, i-Rheo evaluates the complex modulus $G^{*}(\omega) = G\;'(\omega) + iG\;''(\omega)$ from the Fourier transform of the time derivative of $G(t)$ using the piecewise-linear interpolation method originally proposed by Evans et al. \cite{evans2009direct} and enhanced by Tassieri et al. \cite{tassieri2012microrheology} with an oversampling technique. Figure \ref{fr1} shows $G\;'(\omega) $ and $G\;''(\omega) $, computed from the $G(t)$ data in Figure \ref{fg1}, for both explicit-chain BD and active learning-informed Pop-BD. The agreement extends across the accessible frequency range ($10^7$-$10^9$ rad/s), with both methods capturing the characteristic crossover where $G\;'$ equals $G\;''$, corresponding to the primary network relaxation time. The elastic modulus $G\;'$ dominates at high frequencies, reflecting the transient network structure, while $G\;''$ becomes comparable to $G\;'$ at intermediate frequencies as the network relaxes. This frequency-domain comparison provides additional confidence that Pop-BD with active learning metamodels accurately reproduces the viscoelastic response measured in typical oscillatory rheometry experiments. Importantly, our predicted frequency-dependent moduli exhibit deviations from classic terminal behavior at intermediate frequencies. In the terminal regime, viscoelastic liquids typically show $G\;' \sim \omega^2$ and $G\;''  \sim \omega$. However, both our simulations and experimental studies of HEUR-latex suspensions \cite{ginzburg2018oscillatory, pham1999polymeric} reveal weaker frequency dependence with power-law scaling $G\;' \sim G\;''  \sim \omega ^n$ where $n\approx$ 0.6-0.7. This non-terminal behavior, characteristic of nearly-gelled systems with fractal cluster distributions, arises from the broad distribution of relaxation times in these bridged colloidal networks, spanning from rapid bond-level relaxation to slower cluster reorganization timescales. At the highest accessible frequencies ($\omega > 10^9$ rad/s), $G''$ exhibits a pronounced peak followed by a rapid decay, while $G'$ continues to increase and plateaus. This high-frequency behavior reflects the transition from viscous-dominated to elastic-dominated response when the observation timescale (1/$\omega$) becomes shorter than the fastest relaxation processes in the system. Under such rapid loading cycles, the transient network responds predominantly through elastic deformation of polymer chains and bonds rather than through bond breaking and reformation, resulting in minimal energy dissipation and the observed decay in $G''$.

The computational efficiency of Pop-BD becomes evident when comparing wall-clock times for examined simulations. For the system of eight colloidal particles, the explicit-chain BD simulation with 80 polymer chains (888 total particles) required approximately 165 hours ($\sim$7 days) on a single NVIDIA A100 GPU to complete 4$\times 10^9$ timesteps at an average throughput of 6,805 timesteps per second. The corresponding Pop-BD simulation with implicit polymer representation completed the same number of timesteps in approximately 14.5 hours on a single CPU core, achieving an average throughput of 75,000 timesteps per second during the production run. This represents an 11-fold reduction in wall-clock time while using significantly less expensive computational resources. The efficiency gain arises from Pop-BD's coarse-grained representation, which eliminates the need to track individual polymer beads and bond fluctuations, thereby reducing the system from 888 particles to 8 particles with dynamic bonds. This computational advantage becomes even more pronounced for larger systems: simulating 80 colloidal particles with explicit chains to physical times near 1 second would require several months of GPU computation, whereas Pop-BD completes these simulations in days on standard CPUs (Results presented in Figures \ref{f608}–\ref{f121}).

\subsubsection{Active learning-informed Pop-BD simulation results}\label{R3}
In this section, we perform Pop-BD simulations with a focus on the largest relaxation times for relatively larger systems, which are computationally demanding for fully explicit BD simulations. Unless otherwise specified, simulations details are as follows: $\phi = 28\%$, $n_{col} = 80$, $N_k = 20$, $R_{p}$ = 10$b_k$, and $\varepsilon_s = 8 k_B T$, $n_{pol}/n_{col}$=10, which we take as “standard” values. It should be noted that total number of particles is chosen to be 80 to allow sufficient statistics, and it can be increased considering a computational cost of close to $O(N)$ \cite{travitz2021multiscale}, and each simulation was repeated five times with different random seeds, with error bars showing mean values and standard deviations of means.

\begin{figure}
    \centering
    \includegraphics[width=0.4\textwidth]{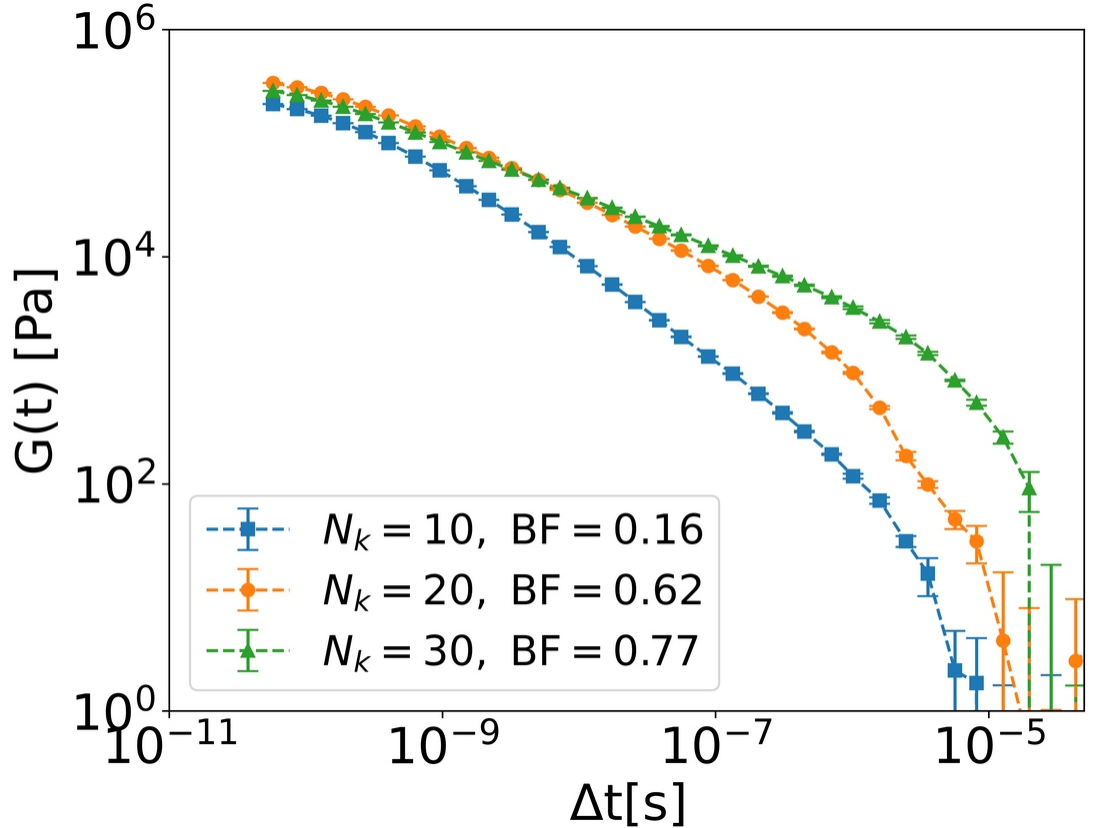}

    \caption{The stress relaxation modulus \( G(t) \) from Pop-BD simulations was analysed for different chain lengths \( N_k \) at $\phi=$ 28\%, involving 80 particles with a radius of \( R_p/b_k = 10 \) and \( n_{pol}/n_{col} = 10 \), and  \( \varepsilon_s = 10 k_B T \). Longer chain lengths (\( N_k \)) lead to higher \( G(t) \) values, indicating a stiffer network with higher number of bridge fractions (BFs) due to improved connectivity among particles. }
    \label{f606}
\end{figure}

Figure \ref{f606} presents the stress relaxation modulus \( G(t) \) from active learning-informed Pop-BD simulations conducted for different chain lengths \( N_k \), while maintaining a constant particle volume fraction \(\phi\) of 28\% and using 80 particles. The other system parameters are set to the standard values mentioned earlier. As evident, longer chains (i.e., higher \( N_k \)) result in higher values of \( G(t) \), indicating a stiffer network of interconnected particles. This behaviour is due to the fact that longer chains have a greater tendency to maintain the network structure, thereby enhancing mechanical rigidity and more prolonged relaxation dynamics. If the number of polymers in a loop configuration at each time step is given by \( n_{loop} = \sum_{i=1}^{n_{col}} N_{ii} \), then the number of polymers in a bridge configuration can be expressed as \( n_{bridge} = n_{pol} - n_{loop} \), where \( n_{pol} \) represents the total number of polymers. The time-averaged bridge fraction, calculated as \( BF = \langle n_{bridge} \rangle / n_{pol} \), indicates the mean fraction of all polymers forming bridges over the course of the simulation and is presented in Figure \ref{f606}. This quantity is determined by counting all bonds (implicit bridges) formed at each dumped time step, averaging these values over all captured configurations during the simulation, and normalizing by the total number of available binding sites (i.e., $n_{pol}/n_{col} \times n_{col}$). A greater \( N_k \) facilitates the formation of a higher number of bridges, which is associated with the enhanced mechanical rigidity and prolonged relaxation dynamics.

\begin{figure}
    \centering
    \includegraphics[width=0.4\textwidth]{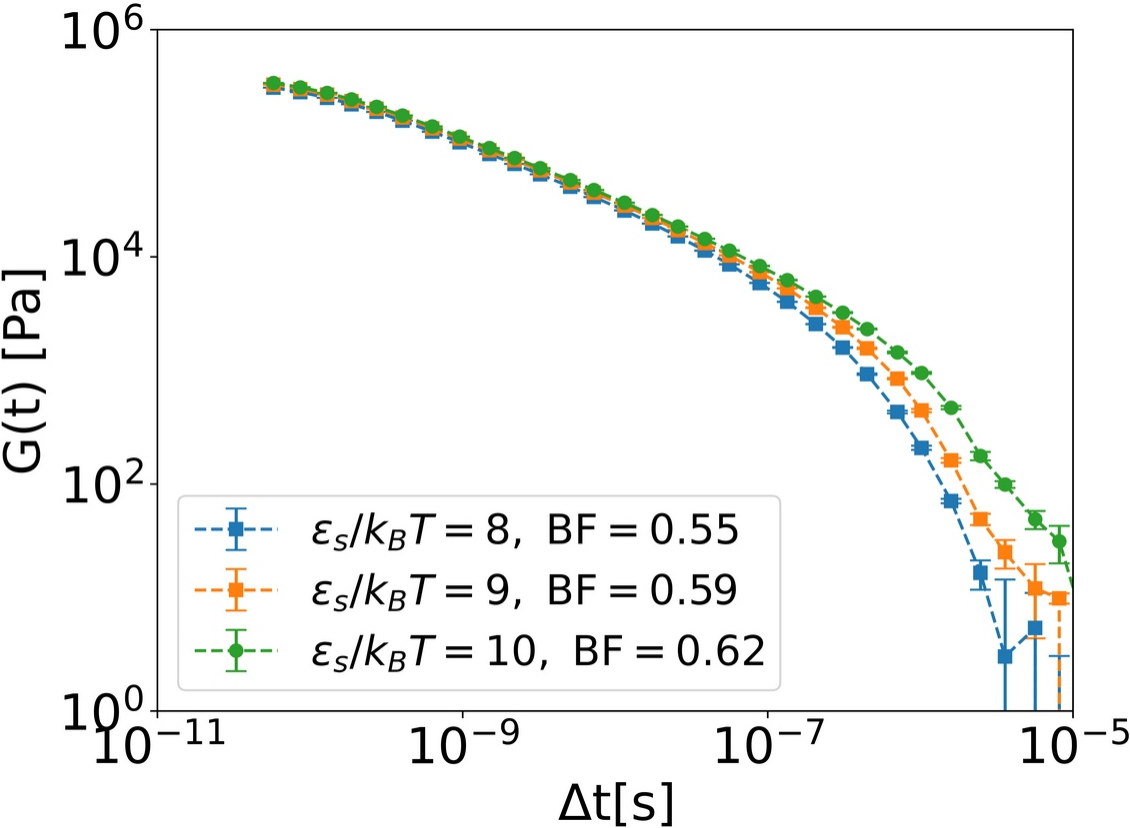}
    \caption{\label{f607} The stress relaxation modulus \( G(t) \) from Pop-BD simulations was examined with varying sticker strengths \( \varepsilon_s \), using \(\phi=\)  28\%,  a total of 80 particles, \( N_k = 20 \), \( R_p/b_k = 10 \), and \( n_{pol}/n_{col} = 10 \). Higher values of \( \varepsilon_s \) resulted in an increased \( G(t) \), a prolonged relaxation time and promoted number of bridge fractions. }
    \end{figure}

Figure \ref{f607} shows the stress relaxation modulus \( G(t) \) obtained from Pop-BD simulations, which were conducted with varying strengths of sticker attraction \( \varepsilon_s \), at constant particle volume fraction \(\phi\)  of 28\% and 80 particles. The other parameters of the system are set as the standard values. It is evident that systems with a stronger sticker attraction or a higher \( \varepsilon_s \) demostrate increased values of \( G(t) \), particularly at extended relaxation dynamics. This occurs because chains with stronger sticker attractions relative to the colloidal particles are better able to maintain the network structure, enhancing mechanical rigidity and leading to extended relaxation dynamics. Additionally, A higher \( \varepsilon_s \) promotes the formation of a greater number of bridges, which correlates with improved mechanical rigidity and prolonged relaxation dynamics, as mentioned in the figure's legend.

\begin{figure}
    \centering
    \includegraphics[width=0.4\textwidth]{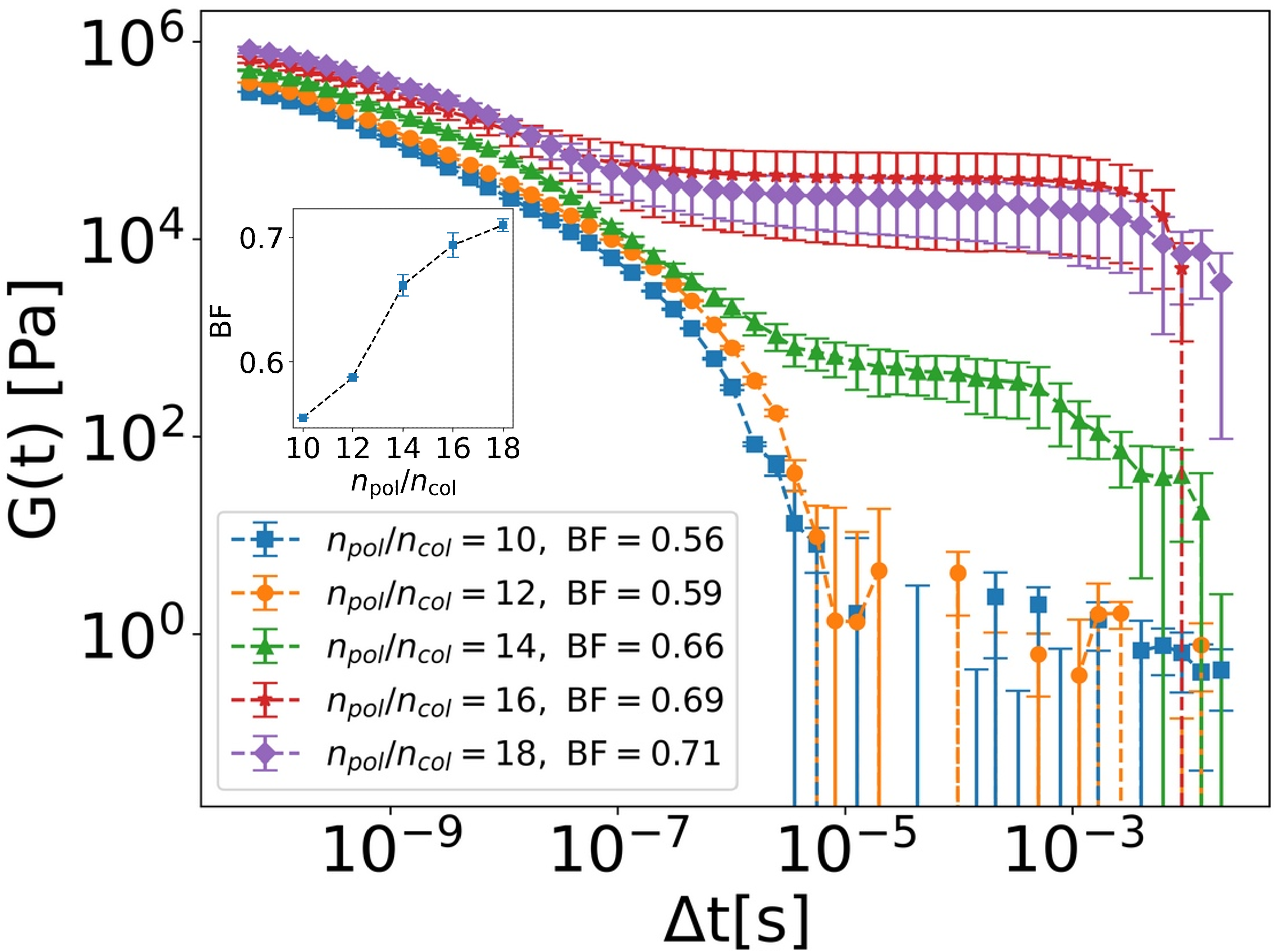}
    \caption{\label{f608} Influence of number of chain-to-particle ratio on stress relaxation and network formation. Stress relaxation modulus $G(t)$ for systems with $n_{pol}/n_{col}$ ranging from 10 to 18, at fixed $\phi = 28\%$, $n_{col} = 80$, $N_k = 20$, $R_p/b_k = 10$, and $\varepsilon_s = 8 k_B T$. Increasing $n_{pol}/n_{col}$ slows relaxation and raises $G(t)$ at long times, indicating stronger network connectivity. The BF increases monotonically, with a distinct jump between ratios 12 and 14, showing a change in the particles networks, as illustrated in the inserted figure.}
    \end{figure}

The ratio of polymer chains to colloidal particles ($n_{pol}/n_{col}$) represents a critical formulation parameter that directly controls the availability of polymers for bridging. We systematically investigated systems with $n_{pol}/n_{col}$ ranging from 10 to 18, keeping other parameter constant as the standard values. Figure \ref{f608} presents both the stress relaxation modulus $G(t)$ and bridge fraction (BF). The stress relaxation data reveal a striking dependence on chain density. At the lowest ratio ($n_{pol}/n_{col}$ = 10), the system exhibits relatively rapid relaxation with $G(t)$ decaying by nearly an order of magnitude over the observation window. As the chain-to-particle ratio increases to 12 and 14, the relaxation becomes progressively slower, with more persistent stress at long times. The highest ratios ($n_{pol}/n_{col}$ = 16 and 18) show markedly extended relaxation, with the modulus remaining elevated beyond $10^{-2}$ s. The bond fraction (BF) defined as the mean value for number of bonds across the simulations run over total possible bonds are also calculated and reported in Figure \ref{f608}. As expected, increasing the total number of chains per particles leads to increased BF monotonically, with a jump from $n_{pol}/n_{col}$ = 12 to $n_{pol}/n_{col}$ = 14, marking the onset of a more interconnected and packed particle networks.

\begin{figure}
    \centering
    \includegraphics[width=0.4\textwidth]{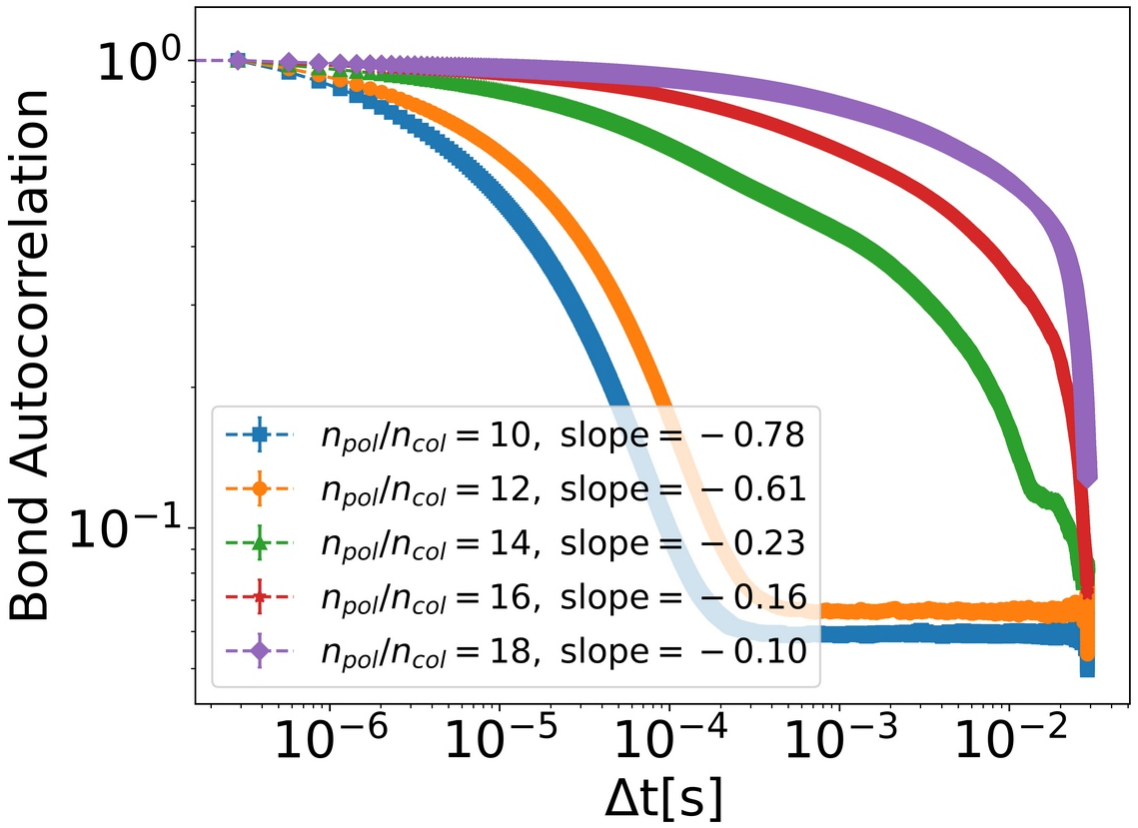}
    \caption{\label{f609}The normalized bond autocorrelation quantifies the persistence of polymer-mediated bridges for systems with $\phi = 28\%$, $n_{col} = 80$, $N_k = 20$, $R_{p}/b_k$ = 10, and $\varepsilon_s = 8 k_B T$. At short lag times, all systems show near-unity correlation, while at longer times, decay becomes progressively slower with increasing $n_{pol}/n_{col}$, indicating more stable and long-lived bridging. A distinct transition in relaxation behavior occurs between $n_{pol}/n_{col} = 12$ and $14$, marking the formation of a more connected colloidal network.}
    \end{figure}

The bond autocorrelation function (BAF) provides quantitative insight into the temporal persistence (memory) of interparticle bridges, capturing how long a given bond between two particles remains intact before breaking. BAF is a function of the lag time, $\Delta t$, and defined as:
\begin{equation}\label{eq_utility_513}
BAF(\Delta t)=\langle N_{ij}(t)\, N_{ij}(t+\Delta t) \rangle,
\end{equation}
where $N_{ij}$ is the number of bonds between particle i and j, and $<>$ denotes ensemble averaging over the simulation duration. For the same system as above($\phi$ = 28\%, $n_{col}$ = 80, $N_k $= 20, $R_{p}$ = 10, and \( \varepsilon_s  \) = 8  $k_B T$), the BAF is normalized by its initial or maximum values ($\langle N_{ij}(t)\, N_{ij}(t) \rangle$) and shown in Figure \ref{f609}. At short lag times ($\Delta t<10^{-6} s$ ), all systems show similar autocorrelation near unity, indicating that bridges remain intact on sub-microsecond timescales. As lag time increases, the autocorrelation decays with distinct slopes that depend on  $n_{pol}/n_{col}$. Systems with lower chain density ( $n_{pol}/n_{col}$ = 10) exhibit steep initial decay (slope = -0.78), indicating rapid bond breaking. With increasing ratios, the slopes progressively decrease, reflecting longer-lived bridges. The  $n_{pol}/n_{col}$ = 18 system displays the shallowest decay (slope = -0.1), consistent with highly stable bridging configurations. At long lag times ($\Delta t>10^{-4} s$), the autocorrelation plateaus at values inversely related to chain density: lower ratios reach near-zero correlation as bridges fully decorrelate, while higher ratios maintain residual correlation reflecting persistent bridging structures. The change in the nature of the network is also noticeable when moving from $n_{pol}/n_{col}$ = 12 to $n_{pol}/n_{col}$ = 14, with a distinct decrease in the rate of fading the bond autocorrelation values. It may seem surprising that increasing the concentration of polymers without modifying their sticker strength or chain length would increase the lifetime of an association. We note that increasing $n_{pol}$ does not significantly change the loop or bridge lifetime for an individual chain, but a multiplicity of bridges between neighbouring colloids would plausibly bind colloids in a network, giving each individual chain time to break and reform a bridge. The net result is the behavior observed in Figure \ref{f609}, that more polymers per colloid yield longer-lasting associations between bound pairs.

\begin{figure}
    \centering
    
    \subfigure[]{
        \includegraphics[width=0.35\textwidth]{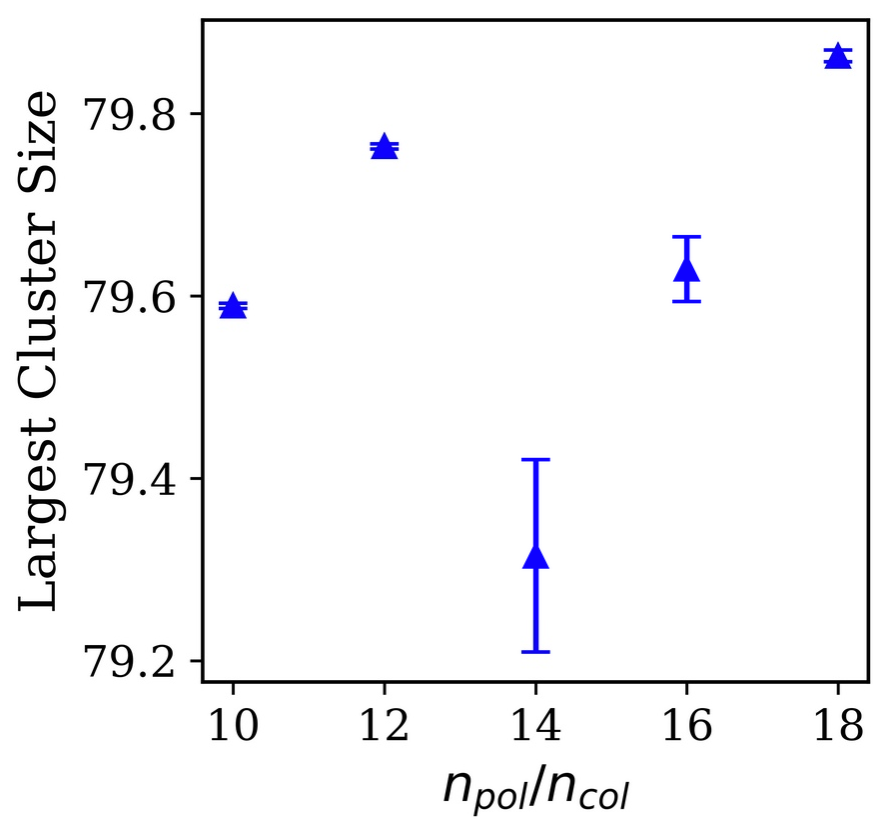}}

    \subfigure[]{
        \includegraphics[width=0.35\textwidth]{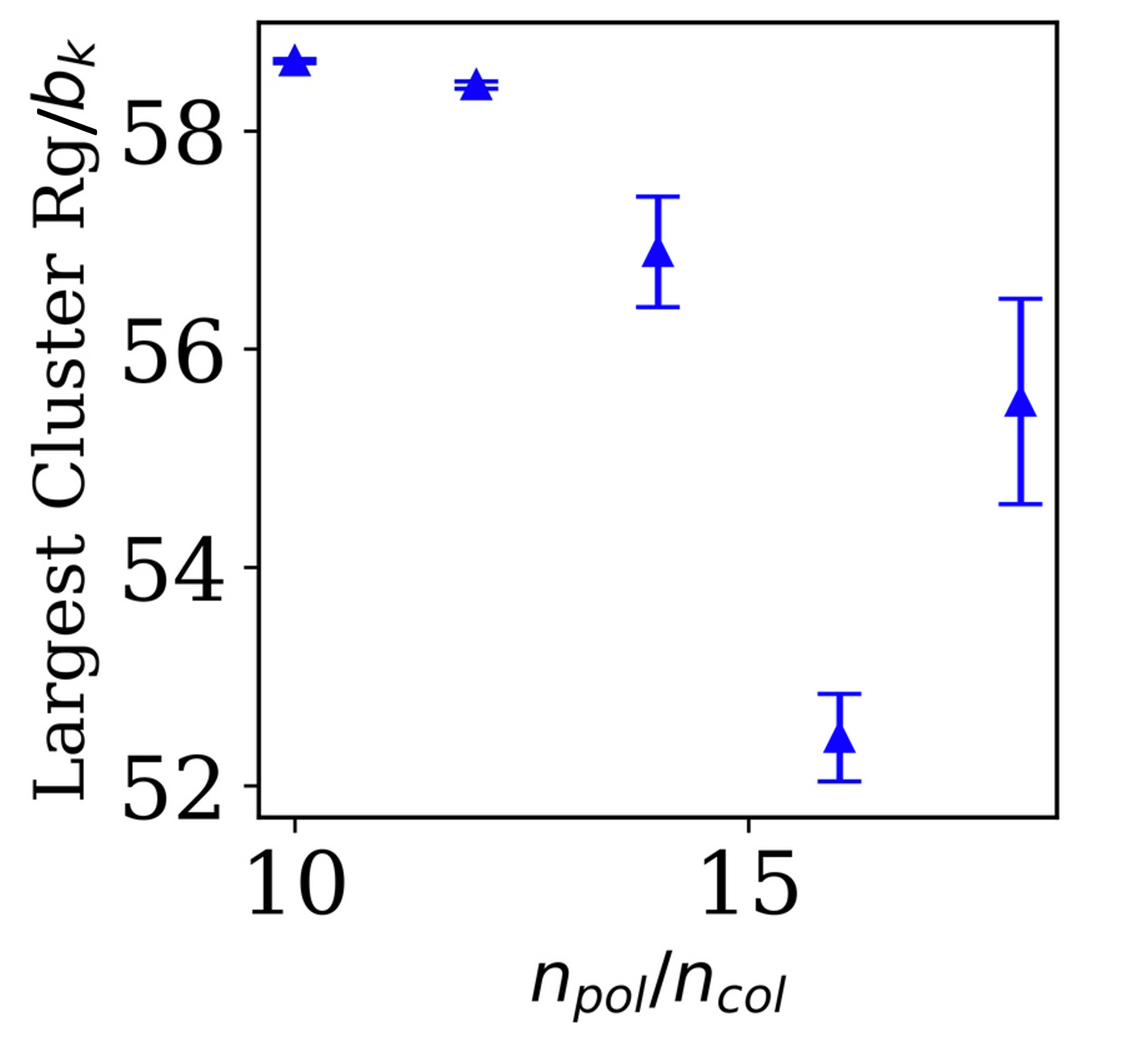}}

    \subfigure[]{
        \includegraphics[width=0.35\textwidth]{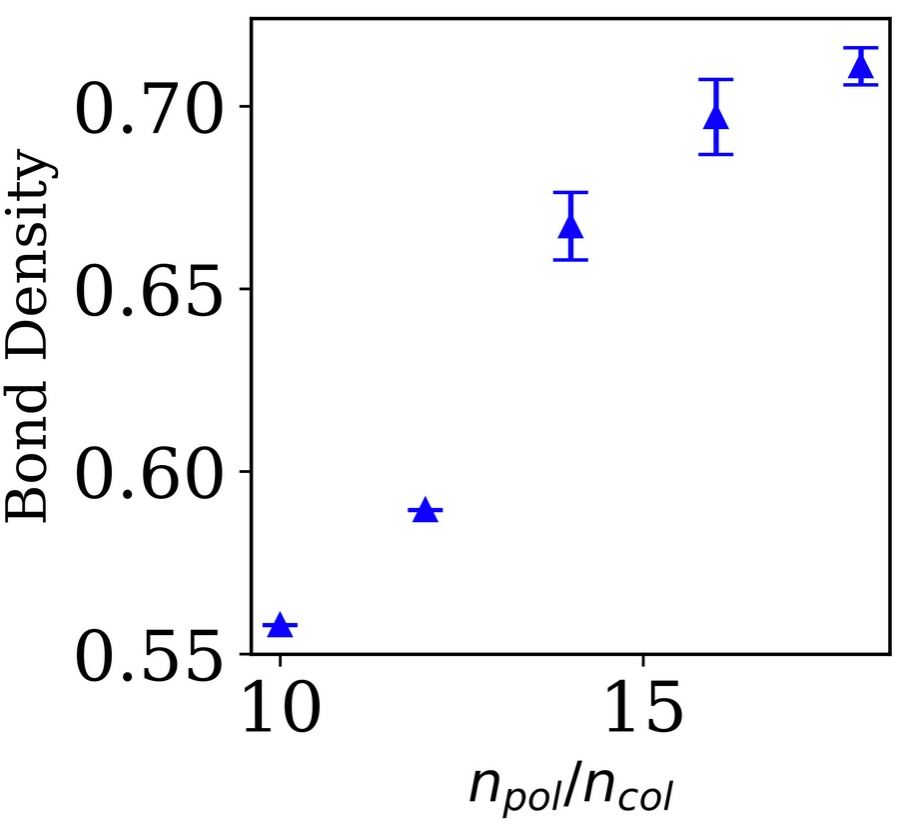}}

    \caption{Cluster analysis of equilibrated Pop-BD configurations at varying chain-to-particle ratios. Particles connected by polymer bridges were grouped using a depth-first search algorithm, and three metrics were evaluated: (a) size of the largest cluster, (b) its radius of gyration $ R_g$, and (c) bond density. Nearly all 80 particles belong to the largest cluster across all $n_{pol}/n_{col}$, indicating a consistently percolated network. Increasing chain loading makes the network more compact, with a slight non-monotonic $ R_g$ suggesting structural fluctuations, while the Bond density rises steadily.}
    \label{f120}
\end{figure}

To gain deeper insight into the network structure underlying the observed rheological behavior, we performed cluster analysis on equilibrated Pop-BD configurations with the same system parameter explored above and varying chain-to-particle ratios, as depicted in Figure \ref{f120}. Clusters were identified using a depth-first search algorithm that connects particles linked by polymer bonds. Three key metrics characterize the network topology: the size of the largest cluster (number of particles), the radius of gyration, $ R_g$, of the largest cluster, and the bond density within the largest cluster. Figure \ref{f120}-a demonstrates that at any ratio of $n_{pol}/n_{col}$, nearly all 80 particles belong to the largest cluster, indicating a percolated network despite the varied polymer content. The radius of gyration, $ R_g$, of the largest cluster reduced, i.e., more compact network as $n_{pol}/n_{col}$ increases as shown in Figure \ref{f120}-b . The non-monotonic $R_g$ trend at increased $n_{pol}/n_{col}$ indicates enhanced structural fluctuations as the system explores various network configurations. There is a certain limitation on how close particles can be to each other, which is analogous to the excluded volume phenomenon. Bond density within the largest cluster quantifies local connectivity, defined as the fraction of polymer chains in bridge configuration within the cluster: 
\begin{equation}\label{eq_utility_bonddensity}
\text{Bond density} = \frac{n_{\text{bridge}}^{\text{cluster}}}{(n_{\text{pol}}/n_{\text{col}}) \times n_{\text{col}}^{\text{cluster}}},
\end{equation}
where $n_{bridge}^{cluster}$ is the number of chains in bridge configuration (counting each bridging chain individually), and $n_{col}^{cluster}$ is the number of particles in the largest cluster. This represents the bridge fraction calculated specifically for chains within the largest cluster. It shows a monotonic increase in bond density directly correlates with the enhanced stress relaxation times and moduli observed in Figure \ref{f608}. 



\begin{figure}
    \centering

    \subfigure[]{
        \includegraphics[width=0.4\textwidth]{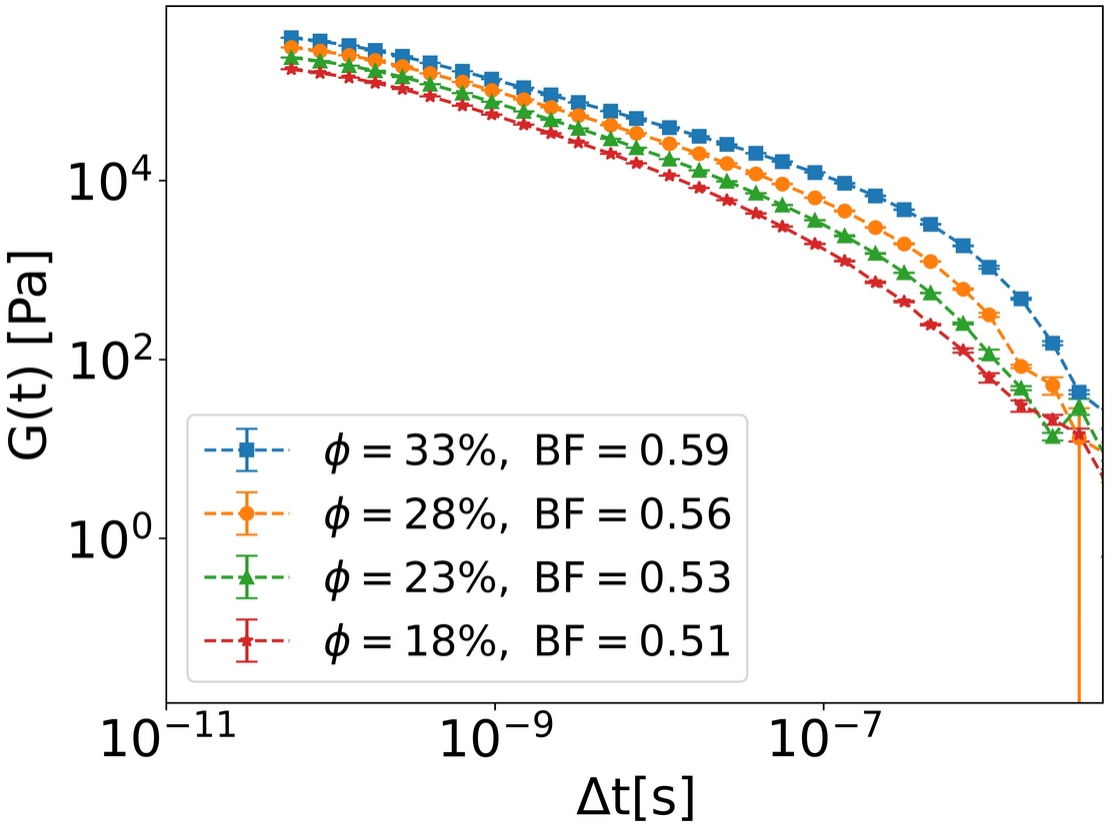}}

    \subfigure[]{
        \includegraphics[width=0.4\textwidth]{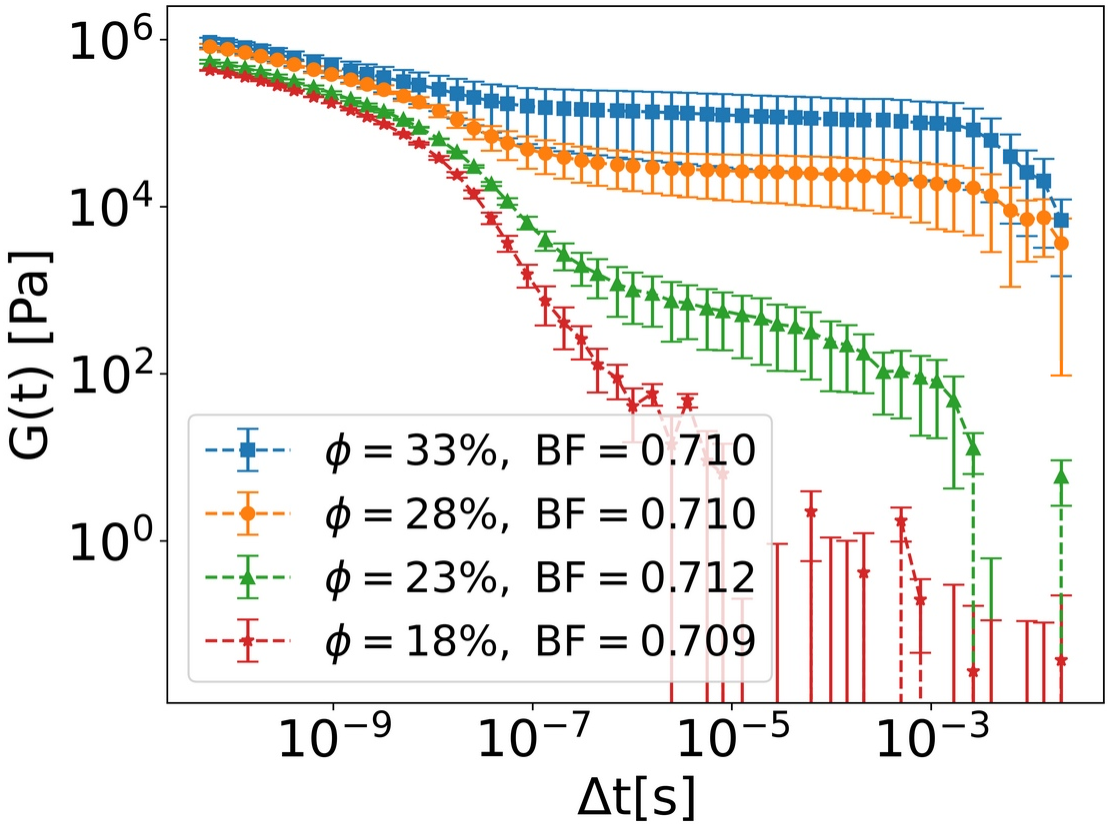}}
    
    \caption{\label{f6111} Influence of particles' volume fraction $\phi$ on stress relaxation and bridge formation. Stress relaxation modulus $G(t)$ for systems with $\phi$ ranging from 18 to 33 $\%$, at  $n_{col} = 80$, $N_k = 20$, $R_{p}/b_k = 10$, and $\varepsilon_s = 8 k_B T$, with polymer-to-colloid ratios of a) $n_{pol}/n_{col}$=10 and b) $n_{pol}/n_{col}$=18. Increasing $\phi$ slows relaxation and raises $G(t)$ at long times, indicating stronger network connectivity. Note that the ranges of the time axes are larger in (b).
 }
    \end{figure}

 The stress relaxation modulus $G(t)$ for different particle volume fractions  ranging from $\phi$ = 18\% to $\phi$ = 33\% are calculated, keeping other parameters constant at $n_{col}$ = 80, $N_k = 20$, $R_{p} = 10$, $\varepsilon_s = 8 k_B T$ , and $n_{pol}/n_{col}$=10 and 18 in Figures \ref{f6111}-a and b, respectively.  Figure \ref{f6111}-a demonstrates that $G(t)$ exhibits two distinct decay regimes.  The initial power-law-like decay ($10^{-8}$ to $10^{-6}$s) corresponds to the breaking of individual bridges, with a range of lifetimes due to varying interparticle gaps.  The slower decay at longer times (> $10^{-6}$ s) reflects topological rearrangements where particles escape from cages formed by neighboring bridged particles \cite{hajizadeh2018novel}.  Higher particle volume fractions yield larger $G(t)$ values across all timescales.  Similarly, the terminal relaxation time increases with particle volume fraction, indicating formation of more persistent particle clusters.  Analysis of the average number of bonds reveals a monotonic increase with increasing $\phi$ as higher particle density both increases the number of neighbors within bridging range and reduces the average gap, promoting bridge formation.  However, for systems with $n_{pol}/n_{col}$=18,  Figure \ref{f6111}-b shows a transition from aforementioned behavior with initial power-law-like decay and network relaxation for  $\phi$ = 18\% to a highly extended network relaxation, manifested by an elevated module beyond $10^{-2}$s for  $\phi$ = 33\%.  Furthermore, the mean bond numbers for the same system are almost the same for the range of $\phi$, showing a saturated system.

\begin{figure}
    \centering

    \subfigure[]{
        \includegraphics[width=0.4\textwidth]{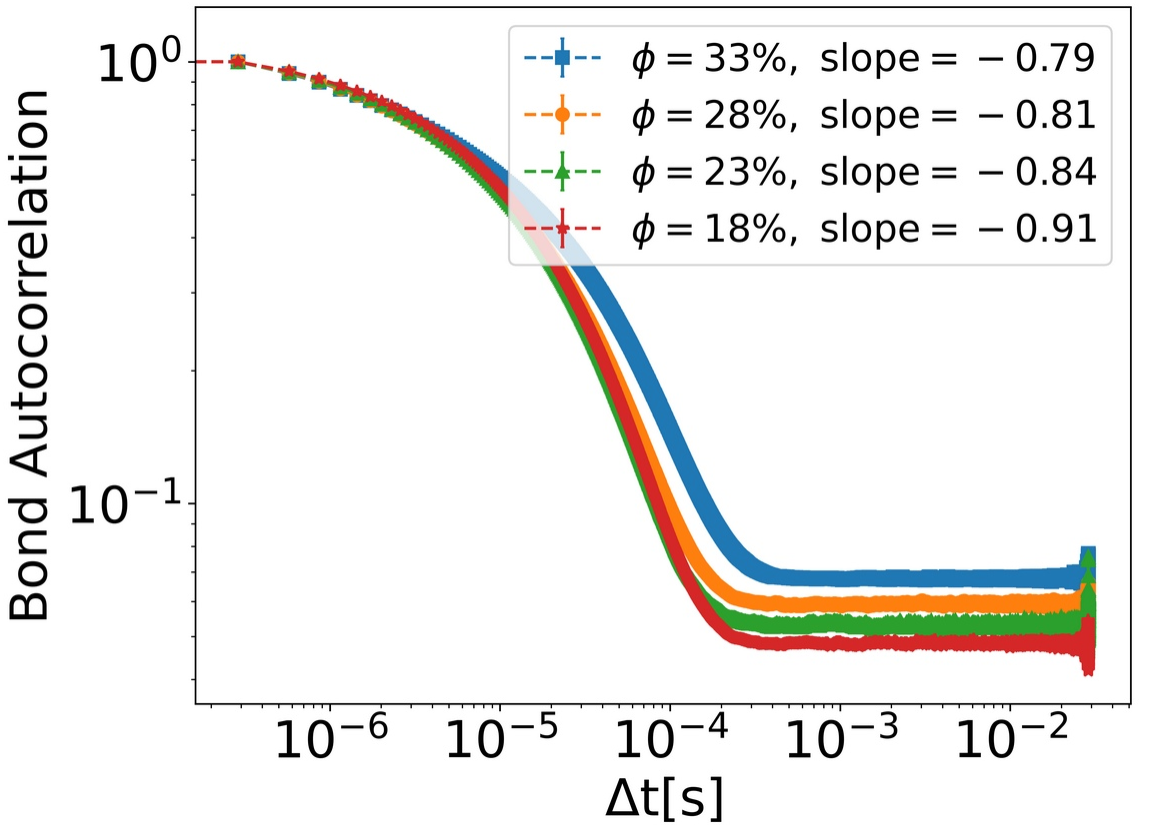}}

    \subfigure[]{
        \includegraphics[width=0.4\textwidth]{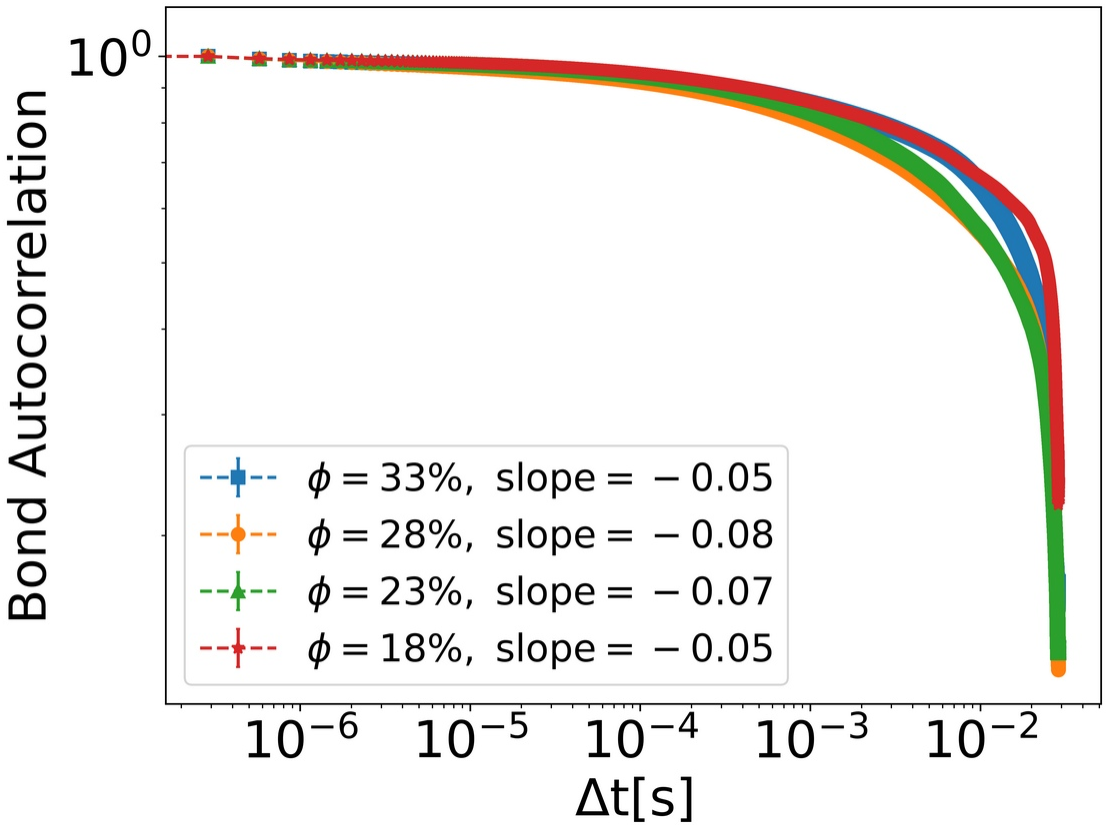}}
    
    \caption{\label{f612}The normalized bond autocorrelation quantifies the persistence of polymer-mediated bridges for systems with $\phi = 28\%$, $n_{col} = 80$, $N_k = 20$, $R_{p}/b_k = 10$, and $\varepsilon_s = 8 k_B T$, at polymer-to-colloid ratios of a) $n_{pol}/n_{col}$=10 and b) $n_{pol}/n_{col}$=18. }
    \end{figure}

\begin{figure}
    \centering

    \subfigure[]{
        \includegraphics[width=0.25\textwidth]{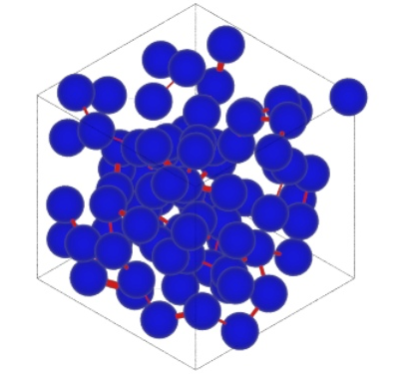}}

    \subfigure[]{
        \includegraphics[width=0.25\textwidth]{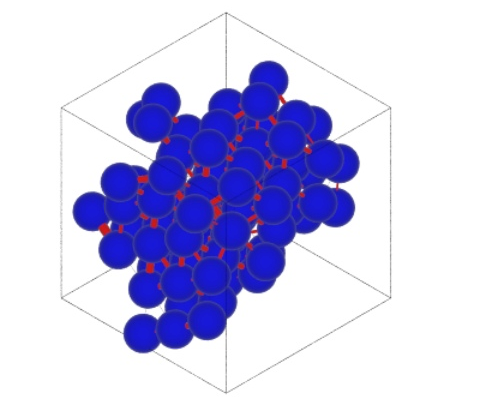}}
    
    \caption{\label{fs}The topological view of the particles at their last snapshots a for systems with $\phi = 28\%$, $n_{col} = 80$, $N_k = 20$, $R_{p}/b_k = 10$, and $\varepsilon_s = 8 k_B T$, at polymer-to-colloid ratios of a) $n_{pol}/n_{col}$=10 and b) $n_{pol}/n_{col}$=18. The red rods represent the bonds between particles, where the thicker ones are associated with higher number of bonds. As evident, the particle network with greater $n_{pol}/n_{col}$ encompasses more bonds and is more compact. }
    \end{figure}

\begin{figure}
    \centering
    
    \subfigure[]{
        \includegraphics[width=0.35\textwidth]{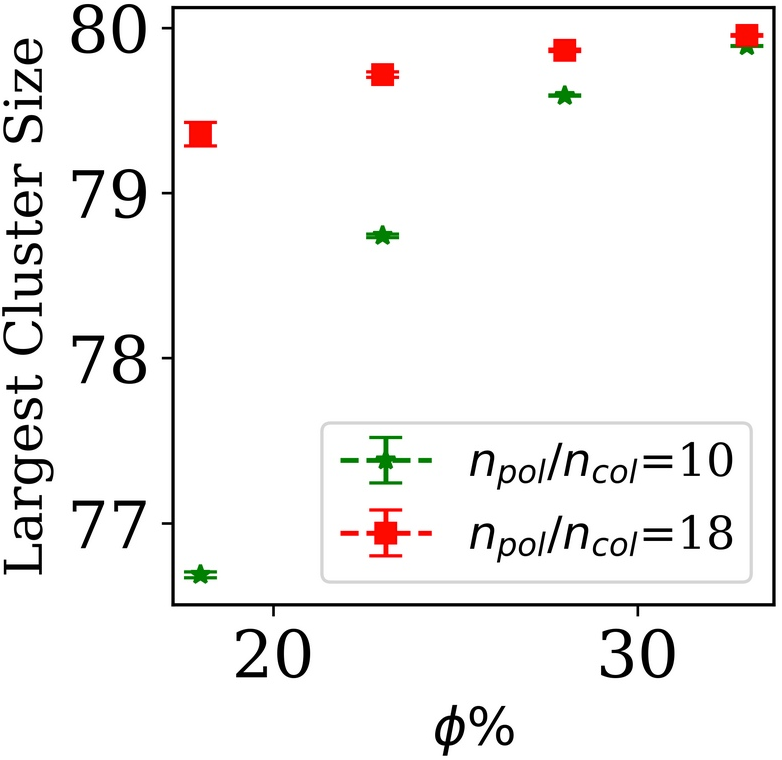}}

    \subfigure[]{
        \includegraphics[width=0.35\textwidth]{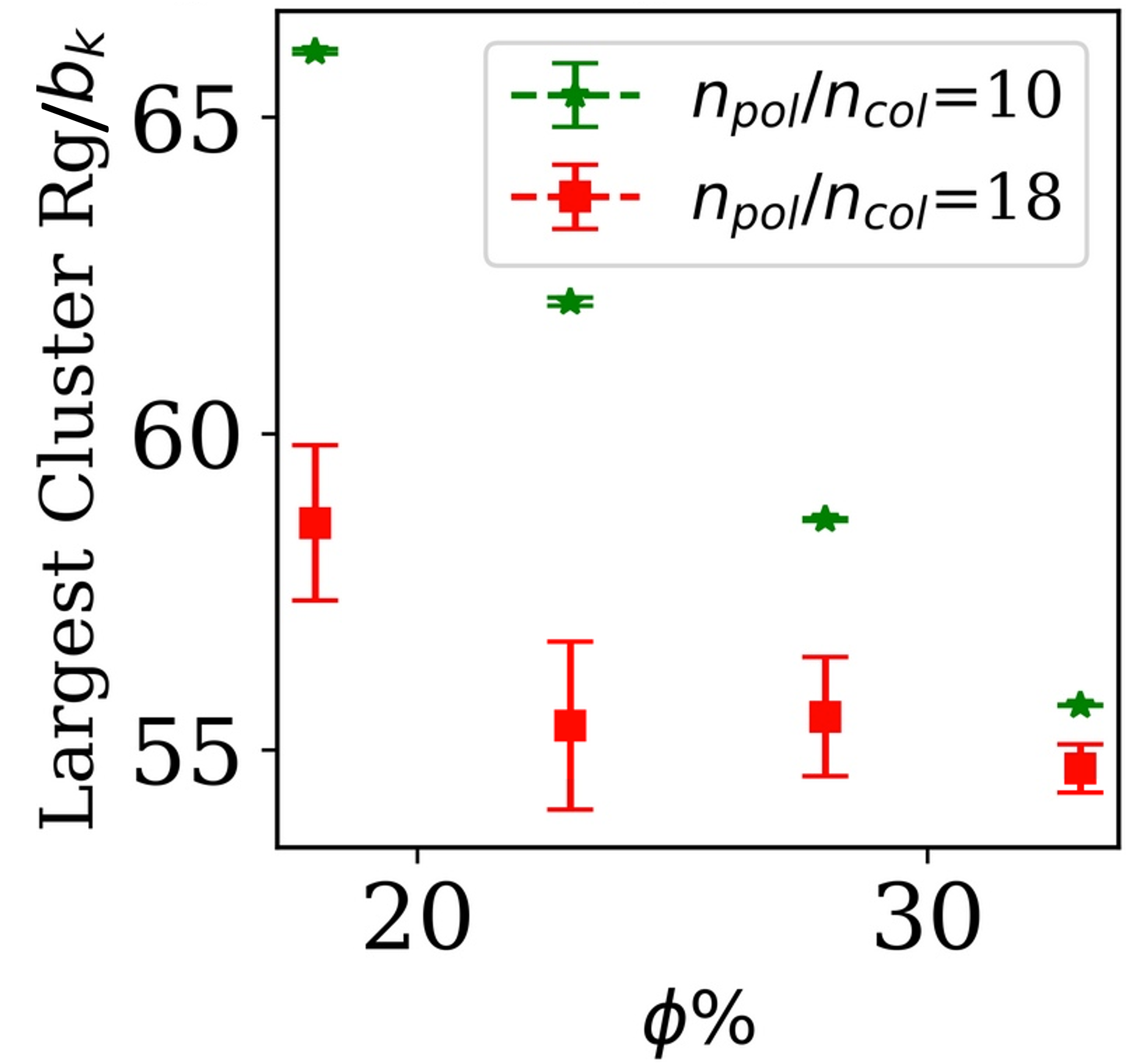}}

    \subfigure[]{
        \includegraphics[width=0.35\textwidth]{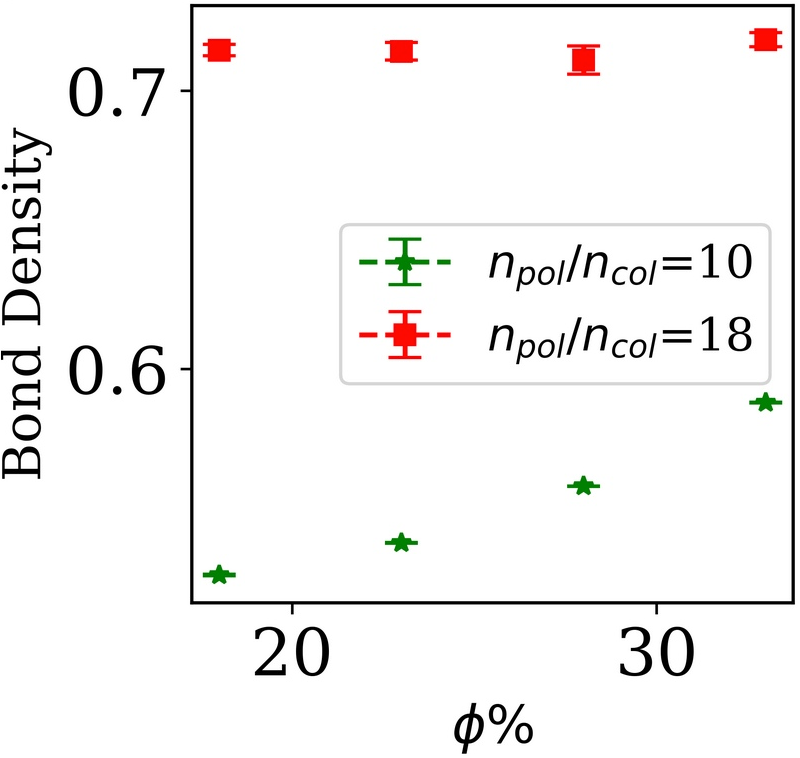}}

    \caption{(a) The largest cluster size increases strongly with particle volume fraction $\phi$, with only \~75\% of particles belonging to the largest cluster at $\phi$=18\% , rising to nearly 100\% at higher $\phi$. (b–c) As $\phi$ increases, the largest cluster becomes more compact and exhibits higher bond density, reflecting progressively stronger particle connectivity.}
    \label{f121}
\end{figure}

The normalized bond autocorrelations, BAF, for the same systems ($n_{col}$ = 80, $N_k $= 20, $R_{p}$ = 10, and \( \varepsilon_s  \) = 8 $k_B T$, and $n_{pol}/n_{col}$=10 and 18) with varied $\phi$ are calculated and depicted in Figure \ref{f612} . As expected, all systems show a bond autocorrelation near unity at lower lag times. As lag time increases, the autocorrelation decays with distinct slopes with small dependency on $\phi$. However, the slop of bond autocorrelation changes significantly when comparing them for distinct systems with  $n_{pol}/n_{col}$=10 and 18. In systems with a lower ratio of $n_{pol}/n_{col}$, as shown in Figure \ref{f612}-a, the bonds last for a shorter duration, approximately $10^{-4}$ seconds. In contrast, in systems with a higher ratio of $n_{pol}/n_{col}$, depicted in Figure \ref{f612}-b, the bonds tend to persist throughout nearly the entire simulation. At extended lag times ($10^{-3}$ to $10^{-1}$ s in Figure \ref{f6111}-a), the normalized BAF exhibits non-zero plateaus that increase with $\phi$. This reflects a finite-size effect where limited particle numbers create a finite probability that particle pairs, at the beginning of the simulation, remain neighbors at later times, even if individual bonds repeatedly break and reform. The increase with $\phi$ arises because particles have more neighbors at higher volume fractions. 
However, in systems with higher \( n_{pol}/n_{col} \), no significant correlation between bond autocorrelation and \( \phi \) is observed. In these cases, the system is fully percolated, meaning that changes in \( \phi \) do not affect the bond density, as illustrated in Figure \ref{f6111}-b. The largest cluster size shown in Figure \ref{f121}-a demonstrates a significant dependence on the volume fraction, $\phi$. For a ratio of \( n_{pol}/n_{col}\) = 10, at a lower volume fraction of $\phi$ = 18$\%$, approximately 75\% of the particles belong to the largest cluster. However, this percentage approaches nearly 100\% at higher particle volume fractions. In contrast, for systems with \( n_{pol}/n_{col} \)= 18, nearly all particles belong to the largest cluster across the examined particle volume fraction range. Figure \ref{fs} depicts a snapshot from the final time step, showing that the particle network structure becomes more compact with increased connectivity when \( n_{pol}/n_{col} \) increases from 10 to 18. Furthermore, as the particle volume fraction increases, the largest cluster becomes more compact with a higher bond density, as illustrated in Figures \ref{f121}-b and \ref{f121}-c. Notably, the changes in the largest cluster size and bond density for \( n_{pol}/n_{col} = 10 \) are more pronounced than those for \( n_{pol}/n_{col} = 18 \).


\section{ Conclusions}\label{con}

This work presents a comprehensive multiscale simulation framework for associative polymer-colloid suspensions, aiming to bridge polymer-level dynamics with macroscopic rheological behavior through active learning-informed Population Balance Brownian Dynamics (Pop-BD). The machine learning metamodels were informed by high-fidelity Brownian Dynamics simulations to capture the transition dynamics of fine-grained polymer chains at a fraction of time and computational resources. The validation studies for relatively small systems demonstrate that Pop-BD simulations informed by active learning metamodels achieve excellent quantitative agreement with explicit-chain Brownian Dynamics across accessible timescales (10$^{-8}$ to 10$^{-5}$ s). The stress relaxation modulus $G(t)$ predicted by our framework captures both the magnitude ($ \sim$ 10 - 10$^{6}$ Pa) and primary relaxation times accurately. Furthermore, frequency-domain validation using i-Rheo conversion confirms that the approach reproduces the characteristic crossover behavior between storage and loss moduli ($G\;'$ and $G\;''$), demonstrating its reliability for predicting viscoelastic responses relevant to oscillatory rheometry experiments. As expected, the implicit polymer representation in Pop-BD cannot capture sub-microsecond relaxations from internal chain modes, resulting in plateau behavior at short times rather than the initial decay observed in explicit simulations. However, this limitation does not compromise predictions of network-scale rheology, which governs application-relevant performance metrics.

Our systematic parameter studies reveal clear structure-property relationships that connect formulation variables to their rheological behavior. Longer polymer chains (higher $N_k$) consistently produce stiffer networks with elevated stress relaxation moduli and extended relaxation times, attributable to both increased bridge formation and enhanced network persistence. Systems with stronger sticker attraction (higher $\varepsilon_s$) similarly exhibit enhanced mechanical rigidity and slower structural relaxation, as stronger polymer-particle interactions stabilize bridging configurations and resist network rearrangement. The chain-to-particle ratio ($n_{pol}/n_{col}$) emerges as a particularly powerful formulation lever: modest increases in polymer loading yield dramatic enhancements in both network connectivity and relaxation timescales, with a distinct transition observed between $n_{pol}/n_{col}$ = 12 and 14 marking the onset of highly interconnected networks for the standard system studied in this paper.

The bond autocorrelation analysis provides quantitative insight into the temporal persistence of polymer-mediated bridges, revealing how formulation parameters control the memory of interparticle connections. At higher chain-to-particle ratios, bonds exhibit markedly slower decorrelation, indicating formation of long-lived bridging structures that resist breakage on experimental timescales. This enhanced bond persistence directly correlates with the observed increases in stress relaxation modulus and terminal relaxation time, establishing bond lifetime as a microscopic predictor of macroscopic viscoelasticity.

Particle volume fraction effects demonstrate the interplay between spatial confinement and bridging thermodynamics. At lower chain densities, increasing particle volume fraction from 18\% to 33\% produces systematic increases in G(t) and bond persistence, as reduced interparticle gaps favor bridge formation and stability. However, at higher chain densities, systems become saturated, where bond fractions plateau and relaxation behavior becomes less sensitive to particle volume fraction, indicating that network connectivity has reached a percolation threshold beyond which additional spatial confinement provides diminishing returns.

Cluster analysis reveals the mesoscale network structures underlying the observed rheological behavior. Nearly all particles belong to the largest cluster across all examined formulations, confirming formation of percolated networks even at modest polymer loadings. As chain density increases, these networks become progressively more compact (decreasing radius of gyration) and more interconnected (increasing bond density within the largest cluster). The bond density metric, measuring local connectivity of bridged particles, shows monotonic increase with chain-to-particle ratio and directly correlates with enhanced stress relaxation times and moduli. This structural validation demonstrates that rheological enhancement stems from increases in both the number of stress-bearing pathways and the energy required for topological rearrangement. With increasing particle volume fraction at fixed lower chain to colloid ratio ($n_{pol}/n_{col} $= 10), cluster analysis shows strong dependence of network formation on spatial packing, with largest cluster size increasing from $\sim$75$\%$ at $\phi$ = 18$\%$ to nearly 100 $\%$ at higher particle volume fractions. Simultaneously, networks become more compact with higher bond density, reflecting progressively stronger particle connectivity as colloidal crowding brings particles within bridging range.

The multiscale framework developed here provides practical guidance for rational design of waterborne coating formulations. The clear relationships between molecular parameters (chain length, sticker strength, polymer loading) and rheological properties (modulus, relaxation time, network structure) enable formulators to systematically adjust performance characteristics. For applications requiring high viscosity and elastic character, such as sag resistance and storage stability, our results,  as a proof of concept, suggest targeting higher chain-to-particle ratios ($n_{pol}/n_{col} $ $\geq$ 14) where network percolation and bond persistence are maximized. The observed transitions in network topology and relaxation behavior at specific parameter combinations can guide selection of optimal formulation windows that balance competing performance requirements across application, storage, and leveling stages.


Future work should extend this framework to non-equilibrium conditions, particularly shear flow, where convective transport competes with thermal diffusion in governing polymer association dynamics. Incorporating shear-dependent transition rates into Pop-BD simulations would enable the prediction of the complete flow curve, from zero-shear viscosity through shear-thinning to high-shear limiting behavior, connecting molecular formulation directly to application-relevant rheology. Additionally, expanding the metamodel parameter space to include larger particle sizes, longer chains, and varied particle size distributions would broaden its applicability to commercial coating formulations. Integration with experimental validation across multiple length scales would further strengthen the predictive capability and establish confidence bounds for formulation design. 

Another limitation affecting the results in this study concerns the the distance component of the Lennard-Jones potential, $\sigma_{pp}$, (in Equation \ref{eq_utility_514}) used in the Pop-BD simulations. The current implementation employs the latest version of  purely repulsive Lennard-Jones (LJ) potential between colloidal particles, which  can underestimate the effective repulsion arising from adsorbed polymer loops on particle surfaces. In reality, these loops create a soft repulsive corona around each particle, generating additional steric forces that maintain greater interparticle separation and prevent premature network percolation or particle aggregation. The effect of loop-mediated repulsion was explored in earlier work by Hajizadeh et al. \cite{hajizadeh2018novel}, where an enhanced repulsive potential accounting for the polymer corona was incorporated into the particle-particle interaction framework. This modification resulted in more realistic particle distributions and delayed percolation transitions compared to bare LJ potentials. However, due to insufficient validation against experimental data and explicit-chain simulations across the full parameter space explored in this study, we could not incorporate this enhanced potential with confidence. The systematic validation required to justify such modifications—including comparison of pair distribution functions, cluster size distributions, and rheological predictions against both experimental measurements and explicit-chain BD benchmarks—remains to be completed.

Looking ahead, a promising extension involves addressing the inverse design problem for HEUR-latex formulations using Metropolis Markov Chain Monte Carlo sampling (MCMC). Rather than predicting rheology \cite{MD1,MD2,MD3,MD7,Robe2025} from known formulations (the current forward model), this approach would specify target rheological properties and use Gaussian process (GP) metamodels as a forward model within MCMC to sample the posterior distribution of formulation parameters that achieve these targets. Critically, this Metropolis approach can reveal multiple solutions, providing formulation scientists flexibility to select routes based on cost or processing constraints. For HEUR-latex coatings, similar multiplicity might manifest as different polymer-particle combinations achieving the same application performance, enabling optimization \cite{Ding2025,Weeratunge2022,Weeratunge2025,aplc,el} of secondary objectives beyond rheological targets. 

In conclusion, this work demonstrates that active learning-enhanced multiscale simulation provides a powerful approach for understanding and predicting the rheological behavior of associative polymer-colloid suspensions. By successfully connecting mesoscaled bond dynamics to near macroscopic viscoelastic properties through computationally efficient coarse-grained simulations, we have established a framework that can guide rational formulation design and accelerate development of next-generation waterborne coatings with tailored performance characteristics. Beyond waterborne coatings, the methodological innovations introduced here provide a template for multiscale modeling of other complex soft matter systems.

\begin{acknowledgments}
This research was supported by The University of Melbourne’s Research Computing Services and the Petascale Campus Initiative. JA acknowledges Melbourne Research scholarship award. Any opinions, findings, and conclusions or recommendations expressed in this material are those of the authors and do not necessarily reflect the views of NSF. 

Data Availability Statement: The data that support the findings of this study are available from the corresponding author upon reasonable request.
\end{acknowledgments}

Conflict of Interest (COI): 
The authors have no conflicts to disclose.

Author Contributions:
JA: Writing – original draft; Formal analysis; Software; Investigation; Methodology (lead). DR: Supervision (supporting); Writing – review and editing (supporting); Methodology (supporting). RGL: Conceptualization (supporting); Writing – review and editing (supporting). EH: Supervision (lead); Conceptualization (lead); Writing – review and editing (lead); Funding acquisition.

\mbox{}
\newpage
\mbox{} 
\textbf{References}
\bibliography{aipsamp}

\end{document}